\documentclass[3p,times]{elsarticle}

\usepackage[T1]{fontenc}
\usepackage[english]{babel}
\usepackage{microtype}
\usepackage{graphicx}
\usepackage{subcaption}
\usepackage{csquotes}
\usepackage{amsmath}
\usepackage{amssymb}
\usepackage{booktabs}
\usepackage{comment}
\usepackage{bm}
\usepackage[section]{placeins}
\usepackage[utf8]{inputenc}
\usepackage{makecell}

\usepackage{parskip}

\usepackage{hyperref}
\hypersetup{
    colorlinks  = true,
    linkcolor   = black,
    citecolor   = black,
    urlcolor    = blue
}

\usepackage{cleveref}
\usepackage{algorithm}
\usepackage{algpseudocode}

\usepackage{titlesec}
\titleformat{\section}{\Large\bfseries}{\thesection}{1em}{}
\titleformat{\subsection}{\large\bfseries}{\thesubsection}{1em}{}
\titleformat{\subsubsection}{\normalsize\bfseries}{\thesubsubsection}{1em}{}

\journal{Journal Name}

\begin{document}

\begin{frontmatter}

\title{Generating synthetic evolution of turbulent flames with an experimental data-based spatiotemporal diffusion model}

\author[nd]{Amrit Tarur\corref{cor1}}
\cortext[cor1]{Corresponding author}
\ead{atarur@nd.edu}

\author[nd]{Shivam Barwey}

\address[nd]{Department of Aerospace and Mechanical Engineering, University of Notre Dame}

\begin{abstract}

In this study, a conditional diffusion model -- a class of generative machine learning models -- is developed to generate synthetic, experimental data-based trajectories of turbulent flames. Generated experimental data corresponds to simultaneous field measurements, namely OH planar laser-induced fluorescence (OH-PLIF) fields and multi-component particle image velocimetry (PIV) fields, for attached and detached flame states in a swirl combustor configuration. This is done using an x-prediction flow matching framework combined with a pixel-based spatiotemporal transformer, which is capable of generating entire spatiotemporal slabs containing synthetic flame evolution at inference time, conditioned on the flame regime. Using this framework, synthetic flames were found to preserve key flame features and statistical consistency across space and time, particularly at the large scales -- deviations at high temporal frequencies and small spatial length scales were found to depend on the time-span of the generated space-time slabs. An extrapolation task of transition synthesis is also conducted, in which the conditional diffusion model is used to synthesize spatiotemporally coherent flame transitions (flame liftoff and reattachment) unseen by the model during training. This was accomplished using a model for the denoising transition velocity that relies on time-varying linear combinations of attached and detached denoising velocities, leading to an approach that (a) allows for control of the generated transition directions and timescales, and (b) retains sample-to-sample variability in the generated transitions in the process. Overall, this study provides a promising pathway for the utilization of experimental data-based generative models as a new means of data exploration in data-sparse environments, complementing both experiments and computational fluid dynamics-based approaches.
\end{abstract}

\begin{keyword}
machine learning \sep turbulent flames \sep laser diagnostics \sep generative modeling \sep diffusion model
\end{keyword}
\end{frontmatter}

\section{Introduction}

In the past decade, modeling strategies based on machine learning (ML) have seen a notable rise in the combustion sciences \cite{ml_combustion}. This is due in part to (a) the rapid advance in ML driven by natural language processing and computer vision fields, from which state-of-the-art model developments can be transferred, (b) the ability to synergize ML models with existing physics-based numerical models, and (c) the fact that ML-based approaches - particularly those based on neural networks - are designed to be data-based. That is, they benefit from ingesting large amounts of data in the model development (training) pipeline, a quality that can be interpreted as bypassing the need for extensive human-in-the-loop model development \cite{venkat_emergingtrends}. This quality is desirable in scenarios where large datasets exhibiting complex physics (such as turbulent combustion processes) are available. Conventionally, this data scaling advantage motivates usage of machine learning in conjunction with high-fidelity simulation-generated data (since it is easy to generate data at scale using numerical solvers), the result of which has led to promising advances in numerical combustion modeling \cite{jon_cnf_2025,chemnode,lapeyre_cnn,detonation_ftc} that can enhance purely physics-based approaches. 

However, alongside simulation-based data sources, advances in experimental combustion diagnostics - particularly, laser-based and high-speed imaging technology \cite{steinberg2023optical,frank2021advances} - also continue to accelerate. The result is an ever-expanding source of experimental datasets that, alongside usage for analysis and simulation validation, provide entirely new pathways for combustion and fluid dynamics model development using machine learning \cite{vinuesa_experiments_ml,ml_combustion,venkat_emergingtrends}. So long as sufficient training data is available, experimental data-based machine learning has a key advantage, in that it offers a direct pathway to constrain models to the complex physical information contained in real-world data sources directly. This has led to the development of experimental ML models serving a variety of purposes in combustion applications, including reduced-order modeling \cite{sarkar2015early,sarkar2015dynamic,shivam_crom_ctm}, measurement transformation/completion models \cite{shivam_proci,parente_info_overlap,sitte2022velocity}, and combustion-based computer vision models for feature identification \cite{johnson_rde} and flow segmentation \cite{vansh_sootfoil}.

Alongside the above applications, \textit{generative} machine learning offers particularly powerful pathways for modeling from the data synthesis standpoint, with ``synthesis'' in this context referring to the generation of data samples that were previously unobserved in the training dataset. In broad terms, generative machine learning (which, as the name implies, involves the training and usage of so-called generative models) refers to the process of (a) sampling from high-dimensional conditional distributions for which no closed-form expressions exist, and (b) using this sampling capability to synthesize new, previously unobserved data for the purposes of achieving some modeling goal of interest \cite{lai2026principlesdiffusionmodels}. Unlike deterministic ML approaches, generative ML models rely on a sampling process: a successful generative model, in this sense, is one that demonstrates data synthesis capability in a way that ensures new samples are indeed unique with respect to training data samples, while obeying statistical consistency with respect to the training data distribution \cite{sohldickstein2015deepunsupervisedlearningusing}. In other words, the data generation capability provided by generative models can be interpreted as an exploratory mechanism for complex, high-dimensional probability distributions \cite{malik_gan} -- when such distributions are characterized by complex experimental data (e.g., simultaneous PLIF/PIV measurements, as considered here), such sampling can lead to valuable avenues for surrogate experimental data generation capability that sits alongside computational fluid dynamics-based approaches \cite{CARREON2023100238}.

To this end, the objective of this work is to develop a generative modeling strategy for synthesizing spatiotemporal, simultaneously measured OH-PLIF and multicomponent PIV data from a premixed swirl combustor configuration \cite{AN2019267} -- in particular, it is shown here how a generative model trained to synthesize spatiotemporal flame dynamics in isolated attached and detached flame regimes can also be used to synthesize flame \textit{transition} data between these states, enabling a pathway for extreme event data generation capability \cite{malik_pecs}. 

Before expanding on the specific contributions of this work, a brief overview of recent efforts in generative modeling in combustion is warranted, as it is now a rapidly evolving research area with a wide range of applications. For example, in Ref.~\cite{pinaki_vae_2025}, a variational autoencoder (VAE) -- an approach that facilitates high-dimensional distribution sampling using reduced latent vector transformations -- is developed to learn compressed molecular representations for accelerated fuel design and exploration. In Ref.~\cite{ates_eai_spray}, a conditional generative adversarial network (GAN) is used to synthesize droplet trajectories in an engine combustor configuration, again through the use of reduced latent vector mappings. Alongside GANs and VAEs, diffusion models \cite{lai2026principlesdiffusionmodels} have recently been used for surrogate modeling \cite{li_pof_2026} and flow reconstruction \cite{wu2025multi,guo2026reconstruction} tasks in reacting flow applications, mirroring the rapid advances in diffusion-based generative modeling observed in non-reacting settings \cite{confild,romit_hypersonic_diffusion,karthik_diffusion_2025,farimani_flow_matching}.

Of particular relevance to the present work is Ref.~\cite{CARREON2023100238}, wherein a conditional GAN is used to synthesize \textit{experimental} OH planar laser-induced fluorescence (PLIF) images in a swirl-stabilized combustor configuration. Specifically, through the usage of a clustering algorithm applied to OH-PLIF fields, macroscopic flow states (i.e., attached and detached flame clusters) are used as conditioning variables to generate diverse sets of instantaneous flame images in the OH-PLIF modality. The study in Ref.~\cite{CARREON2023100238} demonstrates how a genenrative model can be used to synthesize new experimental flame images for a single modality (namely, OH-PLIF), in various macroscopic flame states using a transformation from reduced latent vectors into the high-dimensional distribution of OH-PLIF images. 

The approach in Ref.~\cite{CARREON2023100238} lays the groundwork for the present study, which extends the experimental data-based flame generation strategy through the following contributions. First, in the present work, a generative model is constructed to achieve synthesis of \textit{spatiotemporal} flame dynamics (instead of purely instantaneous flame images) in attached and lifted flame regimes, in a multi-modal rather than a uni-modal setting. More specifically, this work provides a methodology to control the synthesis of \textit{trajectories} (time-evolving sequences of images) in two different macroscopic flame states (attached, lifted) for not only OH-PLIF, but also three-component simultaneously measured PIV fields. This is achieved through an ``x-prediction''-based conditional diffusion model \cite{jit} powered by a transformer that operates in space-time. Second, the model developed in this work operates directly in the space of image pixels (the full set of pixels comprising space-time slabs), eliminating the need to define a latent vector to facilitate sampling \cite{jit}, as is often required by VAEs or GANs. Third, once trained, the model is demonstrated on a flame transition data synthesis application that it was not exposed to in the training stage. Here, the conditional diffusion model -- trained to synthesize attached and lifted flame dynamics in isolation -- is used to synthesize spatiotemporally-coherent flame \textit{transitions} (flame liftoff and reattachment) unseen to the model during training, using a post-training assumption that relies on mixing isolated attached and lifted flame dynamics in the generation process.

The remainder of the paper proceeds as follows. In Sec.~\ref{sec:data}, a description of the training dataset is provided. In Sec.~\ref{sec:methods}, the conditional diffusion modeling framework and transformer architecture are described. Results and concluding remarks are then provided in Sec.~\ref{sec:results} and Sec.~\ref{sec:conclusion}, respectively.

\section{Dataset}
\label{sec:data}

The data used in this study corresponds to simultaneously measured, time-resolved OH planar laser-induced fluorescence (PLIF) and three-component stereoscopic particle image velocimetry (PIV) fields of a turbulent premixed flame in a model gas turbine swirl combustor. This is the same data source used in Refs.~\cite{CARREON2023100238,shivam_crom_ctm}. For completeness, a brief overview of the configuration and a data description is first provided here in Sec.~\ref{sec:description_of_data}. Then, Sec.~\ref{sec:time_slab} describes the procedure by which the raw data are converted into a time-slab representation required to train the transformer-based spatiotemporal diffusion model, which is later described in Sec.~\ref{sec:methods}.

\subsection{Description of Data}
\label{sec:description_of_data}

\begin{figure}[ht]
    \centering
    \includegraphics[width=0.9\linewidth, trim={0 0 0 0}, clip]{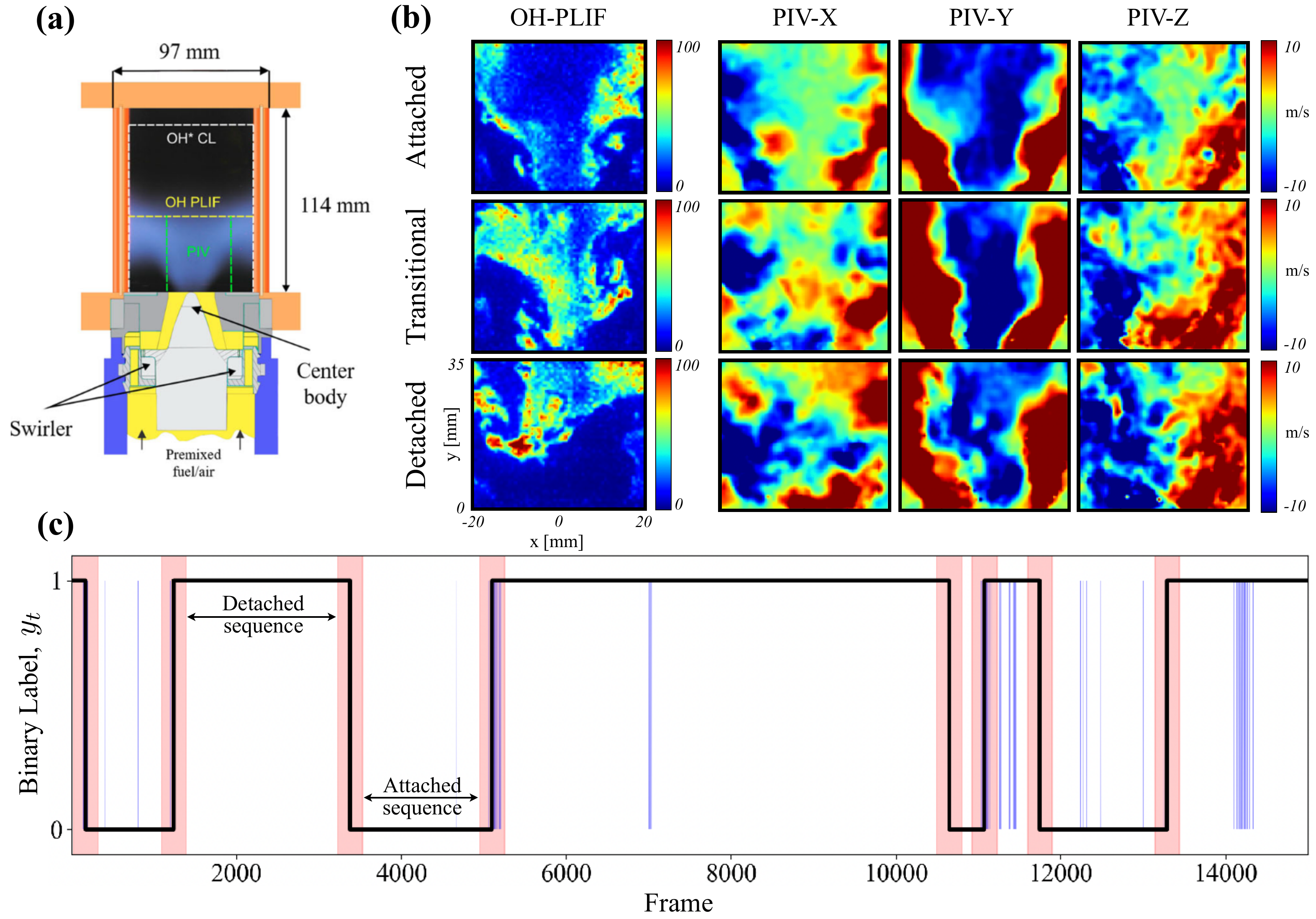}
    \caption{\textbf{(a)} Combustor schematic, showing domain extent of original OH-PLIF and PIV data. \textbf{(b)} Instantaneous flame images after application of pre-processing steps for attached (top row), transitional (during liftoff, middle row), and detached (bottom row) states, showing OH-PLIF and PIV X, Y, and Z modalities in respective columns. \textbf{(c)} Raw binary class label $y_t$ (blue) and the median-filtered signal (black), with transitioning regions excluded from the training dataset highlighted in red. Two illustrative attached and detached sequences are indicated in the image.}
    \label{fig:combustor_dataset}
\end{figure}

The simultaneous OH-PLIF and PIV data come from the experiments of Ref.~\cite{AN2019267}, and the schematic of the gas turbine model combustor is shown in Fig.~\ref{fig:combustor_dataset}(a). Relevant to the present work is the fact that for this swirl combustor, at the operating conditions considered, the flame dynamics admit intermittent and spontaneous transitions between the following two macroscopic turbulent flame states: (1) a V-shaped attached state, characterized by a highly symmetric instantaneous flame structure, and (2) an M-shaped detached state, depicting highly asymmetric instantaneous structure, whose dynamics are anchored to a precessing vortex core (PVC) \cite{AN2016228} that evolves according to a characteristic frequency. As such, both states, while turbulent and complex in their own nature, are dynamically distinct, particularly at large scales. Since there are two macroscopic states, there are two directions by which flame transition can occur: flame liftoff (attached to lifted), and flame reattachment (movement from lifted to attached).

The full dataset comprises 15000 frames, or snapshots, with simultaneous measurements captured at a frequency of 10 kHz, leading to a time step of $10^{-4}$ seconds between snapshots, and a total temporal span of 1.5 seconds. At each time instant, four total spatial fields are available on the same two-dimensional plane: OH radical (from PLIF measurements, with measurements provided in units of relative pixel intensity), and three-component velocity (from PIV measurements, with measurements provided in units of meters per second). The OH-PLIF data has a resolution of $104 \times 62$ pixels, spanning a y-range of $[0.0, 40.48]$~mm and an x-range of $[-32.81, 34.19]$~mm. The PIV data has a resolution of $79 \times 53$ pixels, spanning a smaller y-range of $[0.0, 35.63]$~mm and an x-range of $[-20.18, 33.27]$~mm. As a preprocessing step, due to the difference in domain extents between OH-PLIF and PIV images, all modalities were cropped into an x-range of $[-20,20]$~mm and y-range of $[0,35]$~mm, and were then resampled onto a $128 \times 128$ pixel grid for ease of operation with the transformer approach. Despite the reduced spatial region, this preprocessing step retains the necessary data required to distinguish between attached and lifted flame dynamics, and retains all flame structure complexity. Examples of attached, transitional, and detached flame snapshots after pre-processing are shown in Fig.~\ref{fig:combustor_dataset}(b).

\subsection{Extraction of Spatiotemporal Training Data}
\label{sec:time_slab}

With the full dataset described above, the conversion of this data into a set of training samples required to train the conditional diffusion model is described here. The goal of the conditional diffusion model is to synthesize trajectories (spatiotemporal data) of attached and detached flames. To achieve this, a binary class-conditioning approach is used, with two classes corresponding to attached (integer 0) and detached (integer 1) flame dynamics. At the end of this procedure, a set of training samples are generated. A single training sample corresponds to a spatiotemporal ``slab'' spanning the full spatial domain and a pre-specified time-axis extent (e.g., 100 frames), paired with a conditioning variable that encodes the class for that slab (0 indicating attached, 1 indicating detached). In other words, each training sample is a time slab sourced from either attached or detached sequences, which are continuous-in-time subsets of the full 15000-snapshot trajectory, \textit{excluding the transitions}. As such, training data curation requires two steps: (1) classification of flames into attached, lifted, and transitioning states, and (2) extraction attached and detached space-time slabs from the labels with transitioning states discarded. 

\subsubsection{Classification of flames into attached, lifted, and transitioning states}

Following the procedure in Ref.~\cite{CARREON2023100238}, images are first clustered in the $128\times 128$-dimensional pixel space using K-means \cite{kmeanspp}. The clustering algorithm is applied to the full set of snapshots (using only the OH-PLIF modality), producing a set of clusters and centroids. Each cluster contains a subset of the $15000$ frames, with its centroid serving as an accessible, visualizable conditionally-averaged OH-PLIF field. After clustering ($K=11$ clusters were used here), the subset of clusters corresponding to attached flames were manually identified by inspecting the centroids, and were then used to label each frame as attached or detached as follows: for a given frame $t$, an \textit{attached} label of 0 is assigned if its cluster belongs to the attached set, and a \textit{detached} label of 1 is assigned if its cluster does not belong to the attached set -- the variable $y_t \in [0,1]$ represents the binary class label for frame $t$. 

A plot of $y_t$ versus frame index is shown in Fig.~\ref{fig:combustor_dataset}(c) in blue. As seen in the figure, due to the presence of flame transitions, the raw binary label sequence contains short-duration flickering caused by frames that lie near attached cluster boundaries. Flickering accompanied by a complete and sustained switch in the binary label therefore indicates a flame transition. To detect the specific location of a transition, a median filter with a kernel size of 151 frames is applied to the raw binary labels, serving to suppress the flickers while preserving the true onset and offset of each sustained regime. The filtered signal (the black line in Fig.~\ref{fig:combustor_dataset}), coinciding with an updated class label $y_t$, is then scanned for state transitions, with a buffer of 150 frames on each side of every detected transition point considered \textit{transitional} (the red regions in Fig.~\ref{fig:combustor_dataset}(c)). The transitional frames are then discarded, resulting in a set of transition-free attached and detached \textit{sequences} (two example sequences are indicated in Fig.~\ref{fig:combustor_dataset}(c)), comprised of the full set of frames considered in the training data.

\subsubsection{Extraction of attached and detached spatiotemporal slabs}
Within each sequence, a sliding window of width $T$ frames is advanced with a stride of 1, producing a collection of maximally overlapping spatiotemporal slabs. Each slab is a 4-D tensor $\bm{x}_s \in \mathbb{R}^{C \times T \times H \times W}$, where $C = 4$ corresponds to the four simultaneous measurement channels (OH-PLIF, PIV-$X$, $Y$, $Z$), $T$ is the slab temporal span, $H=W=128$ is the number of pixels along the y- and x-directions, and the $s$ subscript denotes an individual sample (slab) from the training dataset.

Slabs are sampled randomly from the sustained attached and detached sequences during training, with each slab inheriting the regime label of the sequence from which it is drawn -- in other words, the label of the slab, denoted $y_s \in [0,1]$, is extracted from the binary sequence label described above, with $0$ for attached and $1$ for detached. Temporal continuity within a slab is strictly enforced such that no slab is permitted to span more than one sequence.

Two time spans are considered: $T=10$ and $T=100$. At the end of this procedure, the result is a training dataset of $S=12511$ slab-label pairs $\{({\bm x}_s, y_s) \}_{s=1,\ldots,S}$, with 3525 attached and 8986 detached, obtained after discarding transitional frames and using a temporal span of $T=10$. Similarly, the $T=100$ configuration utilizes a dataset of $S=11785$ pairs, comprising 3165 attached and 8620 detached samples.

It is emphasized that through the description of the training dataset above, the generative model is trained purely on the dynamics associated with attached and lifted flames in isolation, with macroscopic transitions between these states held out. The elimination of flame transitions was done for the following reasons: (1) to assess the conditional spatiotemporal generation capability of the model on experimental flame data in fundamentally different dynamical regimes, and (2) to study the capability of applying the trained diffusion model in a truly regime-extrapolative, or out-of-distribution, prediction context; that is, to use the diffusion model to synthesize new flame transitions (both liftoff and reattachment). The latter constitutes a challenging demonstration of generating synthetic data for highly data-sparse experimental regimes (i.e., combustion extreme events) \cite{malik_pecs}.

\section{Methodology}
\label{sec:methods}
With the training dataset described in Sec.~\ref{sec:data}, the method that enables generation of synthetic spatiotemporal flame data is presented as follows. First, in Sec.~\ref{sec:diffusion}, the conditional diffusion modeling framework is presented, which describes the way in which models are trained and used synthesize flame dynamics in attached and detached regimes. In Sec.~\ref{sec:transformer}, the spatiotemporal transformer architecture that is used within the diffusion modeling framework is presented. 

\subsection{Conditional Diffusion Model Framework}

\label{sec:diffusion}

Diffusion modeling is a generative modeling paradigm inspired by thermodynamic principles, originally introduced in Ref.~\cite{sohldickstein2015deepunsupervisedlearningusing}, which has gained immense popularity in recent years over VAE- and GAN-based alternatives due to demonstrated training stability and data scalability. The unifying principle is related to learning, using a neural network, how to ``undo'' a noising process that progressively transforms clean data samples, which are distributed according to a PDF that is unknown, high-dimensional, and cannot be analytically sampled, into standard noise, which is distributed according to a simple standard normal PDF that can be easily sampled. For additional detail on diffusion models and their various interpretations, the reader is directed to Ref.~\cite{lai2026principlesdiffusionmodels}. The diffusion modeling approach used here follows the flow matching formulation of Ref.~\cite{jit}, and is presented below in the context of the flame data described in Sec.~\ref{sec:data}.  

Here, a spatiotemporal flame slab, indicated by the sample ${\bm x}_s$, resides in a phase space of dimension $TCHW$. The collection of all spatiotemporal slabs in the training dataset can then be interpreted as a point cloud in this phase space, with the points distributed according to an unknown data distribution $p_d$ (that is, $\bm{x}_s \sim p_d({\bm x}_s)$). The objective is to produce a sampling approach for $p_d$ such that new synthetic flame slabs in attached and lifted regimes can be feasibly generated.

\begin{figure}[htbp]
    \centering
    \includegraphics[width=0.9\textwidth]{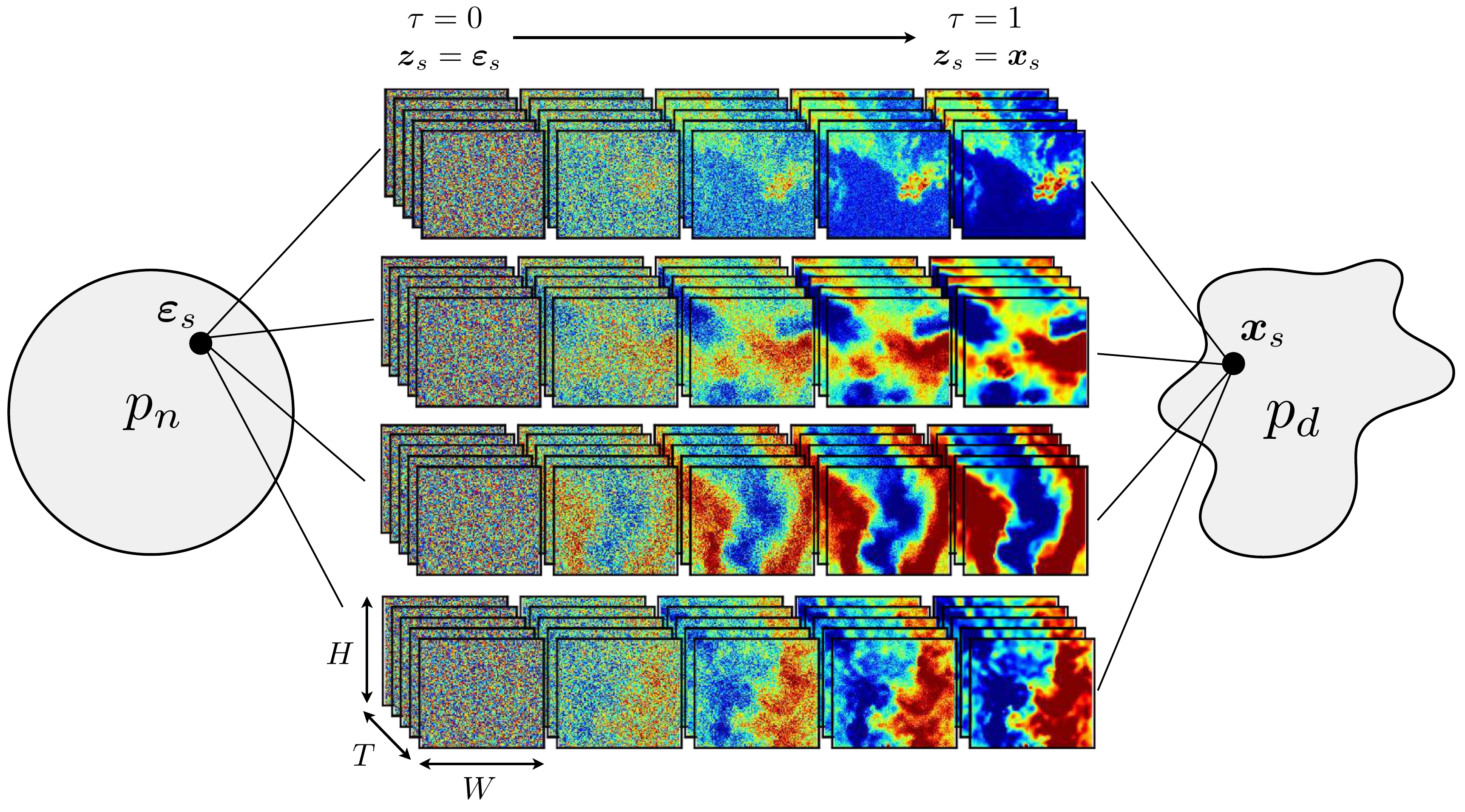}
    \caption{Schematic of the spatiotemporal denoising process for a detached flame, wherein a noise slab $\bm{\varepsilon}_s$ distributed according to the noise distribution $p_n$ is evolved into a flame slab ${\bm x}_s$ distributed according to the data distribution $p_d$. The columns show the evolution of the intermediary noised slab ${\bm z}_s$, with the four measurement modalities (OH-PLIF, PIV-X, Y, and Z respectively) separated by row.}
    \label{fig:Denoise}
\end{figure}

To this end, alongside the data distribution, the diffusion model considers a noise distribution ${\bm \varepsilon}_s \sim p_n({\bm \varepsilon}_s)$, where $p_n = {\cal N}(0, {\bf I})$ is the standard normal distribution in the $TCHW$-dimensional phase space. As such, ${\bm \varepsilon_s}$ is a ``noise slab'' of the same shape as ${\bm x}_s$. The flame slab sample ${\bm x}_s$ arises from a so-called \textit{denoising} process acting on the noise slab ${\bm \varepsilon}_s$. The denoising process occurs along a pseudo-time axis spanning the range $\tau \in [0,1]$, where $\tau$ is the pseudo-time variable, and follows the linear noise schedule
\begin{equation}
    \label{eq:noise_schedule}
    {\bm z}_s(\tau) = \tau\,\bm{x}_s + (1 - \tau)\,\bm{\varepsilon}_s,
\end{equation}
where ${\bm z}_s(\tau)$ is an intermediary noised slab at a point in the $TCHW$-dimensional phase space along the pseudo-time axis between the pure-noise sample ${\bm \varepsilon}_s$ ($\tau = 0$) and the spatiotemporal flame sample ${\bm x}_s$ ($\tau = 1$). A schematic of the denoising procedure is shown in Fig.~\ref{fig:Denoise} -- given the linear noise schedule, sampling is cast as a dynamical evolution from an initial condition at $\tau = 0$ of a pure noise slab evolving along a straight-line trajectory in the phase space to the terminal condition at $\tau = 1$, upon which the intermediary slab ${\bm z}_s(\tau)$ becomes the spatiotemporal flame realization ${\bm x}_s$. 

From Eq.~\ref{eq:noise_schedule}, the rate of change of ${\bm z}_s$ along this straight trajectory is given by
\begin{equation}
    \label{eq:ode}
    \bm{v}_s = \frac{d {\bm z}_s(\tau)}{d\tau} = \mathbf{x}_s - \boldsymbol{\varepsilon}_s. 
\end{equation}
In Eq.~\ref{eq:ode}, ${\bm v}_s$ is the \textit{ground-truth} velocity (or flow -- not to be confused with velocity in the sense of PIV) in the $TCHW$-dimensional phase space required to evolve the noise slab $\bm{\varepsilon}_s$ into the flame slab ${\bm x}_s$. The velocity ${\bm v}_s$ is termed the ground-truth because when sampling in the inference stage, one needs to solve the ODE in Eq.~\ref{eq:ode}; however, one does not have access to the final condition ${\bm x}_s$ required to compute the velocity. As such, the process of diffusion modeling involves constructing a model for the velocity ${\bm v}_s$ such that the ODE in Eq.~\ref{eq:ode} can be solved using a given initial condition ${\bm{\varepsilon}_s}$. With such a model available, for any new realization of ${\bm{\varepsilon}_s}$, an approximation to the ODE in Eq.~\ref{eq:ode} can then be integrated (replacing the ground-truth velocity with the modeled counterpart) to produce a new realization of ${\bm x}_s$, resulting in a sampling process for the flame slabs. A \textit{conditional} diffusion model is obtained by incorporating a conditioning variable (i.e., the desired class label $y_s$ for the sampled flame slab ${\bm x}_s$) into the functional form of the model for the velocity. More formally, 
\begin{equation}
    \label{eq:x_pred}
    \hat{\bm v}_s({\bm z}_s, \tau, y_s) = \hat{\bm x}_s({\bm z}_s, \tau, y_s) - {\bm \varepsilon}_s = f({\bm z}_s, \tau, y_s ; \theta_f) - {\bm \varepsilon}_s , \quad \hat{\bm x}_s = f({\bm z}_s, \tau, y_s ; \theta_f), 
\end{equation}
where $\hat{\bm v}_s$ is the modeled velocity, evaluated using an approximation of the flame slab $\hat{\bm x}_s$ generated by a neural network $f({\bm z}_s, \tau, y_s ; \theta_f)$ -- the neural network, with parameters indicated by $\theta_f$, takes as input the noised slab ${\bm z}_s$, the pseudo-time $\tau$, and the desired class label for the sample $y_s$. The neural network used here is a spatiotemporal transformer, and is discussed further below in Sec.~\ref{sec:transformer}. The formulation in Eq.~\ref{eq:x_pred} is referred to as the ``x-prediction'' variant of diffusion modeling \cite{lai2026principlesdiffusionmodels}, wherein the neural network directly approximates ${\bm x}_s$ to produce the modeled velocity $\hat{\bm v}_s$. While other variants are possible (e.g., v- and $\varepsilon$-prediction), x-prediction is used here due to its reduced sensitivity to input data dimensionality \cite{jit}, which alleviates the need to perform denoising in a reduced latent space and eliminates the associated hyperparameter (latent vector size) for the data manifold dimension.

\textbf{Training and Inference:} The diffusion model is trained by optimizing the parameters of the neural network $f$ via minimization of the mean-squared error (MSE) between predicted and true velocities, resulting in the loss function
\begin{equation}
    \label{eq:loss}
    \mathcal{L}(\theta) = \mathbb{E}_{\bm{x},\,\boldsymbol{\varepsilon},\,\tau}\left[\left\|\hat{\bm v}_s({\bm z}_s, \tau, y_s) - \bm{v}_s\right\|_2^2\right].
\end{equation}
In Eq.~\ref{eq:loss}, the expectation is over a batch of target flame slabs ${\bm x}_s$, noise slabs ${\bm \varepsilon}_s$, and pseudo-time samples $\tau$ drawn from a logit-normal distribution \cite{logit_norm_paper}, with mean $-2.0$ and standard deviation $1.5$. At each training iteration, the intermediary slab ${\bm z}_s(\tau)$ is constructed via Eq.~\ref{eq:noise_schedule}, the ground-truth velocity ${\bm v}_s$ is computed via Eq.~\ref{eq:ode}, and modeled velocity $\hat{\bm v}_s$ is obtained from the x-prediction as 
\begin{equation}
    \label{eq:vpred}
    \hat{\bm{v}}_s({\bm z}_s, \tau, y_s) = \frac{\hat{\bm{x}}_s - {\bm z}_s}{1 - \tau}, 
\end{equation}
which comes from rearrangement per the linear noise schedule in Eq.~\ref{eq:noise_schedule}. The neural network parameters $\theta$ are then updated to minimize $\mathcal{L}(\theta)$ via standard backpropagation-based optimization. The training algorithm is provided in Algorithm~\ref{alg:training}.  

Once the model is trained, inference proceeds by sampling a noise slab ${\bm \varepsilon}_s \sim \mathcal{N}(0, \mathbf{I})$, specifying the attached or detached conditioning label ($y_s \in [0, 1]$), and integrating the ODE
\begin{equation}
    \label{eq:inference_ode}
    \frac{d{\bm z}_s(\tau)}{d\tau} = \hat{\bm v}_s({\bm z}_s, \tau, y_s)
\end{equation}
from $\tau = 0$ to $\tau = 1$. The terminal state ${\bm z}_s(1) = {\bm x}_s$ is the generated spatiotemporal flame slab in the specified class (attached or detached). The inference algorithm is provided in Algorithm~\ref{alg:inference} using forward Euler time integration for simplicity. In this work, the second-order Heun method is utilized in practice for time integration \cite{jit} using 100 integration steps.

\begin{algorithm}[ht]
\caption{Conditional diffusion model training}
\label{alg:training}
\begin{algorithmic}[1]
\Require Dataset $\{(\bm{x}_s, y_s)\}$, neural network $f$, epochs $E$
\For{epoch $= 1, \ldots, E$}
    \For{each mini-batch $(\bm{x}_s, y_s)$}
        \State Sample noise slabs: $\bm{\varepsilon}_s \sim \mathcal{N}(0, \mathbf{I})$
        \State Sample pseudo-times: $\tau \leftarrow \mathrm{LogitNormal}$
        \State Evaluate noised slab: $\bm{z}_s \leftarrow$ Eq.~\ref{eq:noise_schedule}
        \State Evaluate neural network: $\hat{\bm{x}}_s \leftarrow f(\bm{z}_s, \tau, y_s;\theta_f)$
        \State Evaluate modeled velocity: $\hat{\bm{v}}_s \leftarrow $ Eq.~\ref{eq:vpred}
        \State Evaluate target velocity: $\bm{v}_s \leftarrow $ Eq.~\ref{eq:ode}
        \State Evaluate loss: ${\cal L}(\theta_f) \leftarrow$ Eq.~\ref{eq:loss}
        \State Backpropagate and parameter update
    \EndFor
\EndFor
\State \Return trained neural network $f$
\end{algorithmic}
\end{algorithm}

\begin{algorithm}[ht]
\caption{Conditional diffusion model inference}
\label{alg:inference}
\begin{algorithmic}[1]
\Require Trained neural network $f$, class label $y_s \in [0,1]$, integration steps $N$
\State Initialize noise slab: $\bm{z}_s(\tau_0) \sim \mathcal{N}(0, \mathbf{I})$
\State Set pseudo-time grid: $\{\tau_k\}_{k=0}^{N} \leftarrow \mathrm{linspace}(0, 1, N+1)$, \;\; $\Delta\tau \leftarrow 1/N$
\For{$k = 0, 1, \ldots, N-1$}
    \State Evaluate neural network: $\hat{\bm{x}}_s \leftarrow f(\bm{z}_s(\tau_k),\, \tau_k,\, y_s;\theta_f)$
    \State Evaluate modeled velocity: $\hat{\bm{v}}_s \leftarrow$ Eq.~\ref{eq:vpred}
    \State Euler update: $\bm{z}_s(\tau_{k+1}) \leftarrow \bm{z}_s(\tau_k) + \Delta\tau\,\hat{\bm{v}}_s$
\EndFor
\State \Return generated slab $\bm{x}_s \leftarrow \bm{z}_s(\tau_N)$
\end{algorithmic}
\end{algorithm}

\subsection{Description of the Spatiotemporal Transformer}
\label{sec:transformer}

The diffusion modeling framework outlined in Sec.~\ref{sec:diffusion} requires specification of the neural network $f$, which here is chosen to be a spatiotemporal transformer-based architecture that operates directly in the pixel space. Specifically, the model is a spatiotemporal adaptation of the image-based architecture used in Ref.~\cite{jit}, which results in a Diffusion Transformer \cite{dit} operating on patches of pixels in space-time. The transformer approach is especially appealing for the flame generation task due to its ability to (a) model complex non-local interactions across the entire time slab during the generation process, achieved through space-time tokenization combined with multi-head self-attention layers \cite{vaswani}, (b) control the limiting spatiotemporal scales of these modeled interactions via a patch (token) size, and (c) operate directly in pixel space without requiring the specification of a latent dimensionality for the data or noise variables (the latent vector hyperparameter is instead exchanged with a patch size hyperparameter).

The model evaluation is decomposed into the following four steps: (1) tokenization, (2) token and conditioning-variable embedding, (3) the transformer block, and (4) token decoding. Details on each component are provided sequentially below. In the following text, subscripts on the input slab are dropped for notational clarity (${\bm z}_s$ becomes ${\bm z}$, still indicating a single noised flame slab). Before proceeding, it is emphasized that there are three inputs to the model, namely the input noised slab ${\bm z}$, the class label for the slab $y$, and the pseudo-time for the slab $\tau$. The output is $\hat{\bm x}$, which is the predicted denoised slab (following x-prediction).

\begin{figure}
    \centering
    \includegraphics[width=\linewidth]{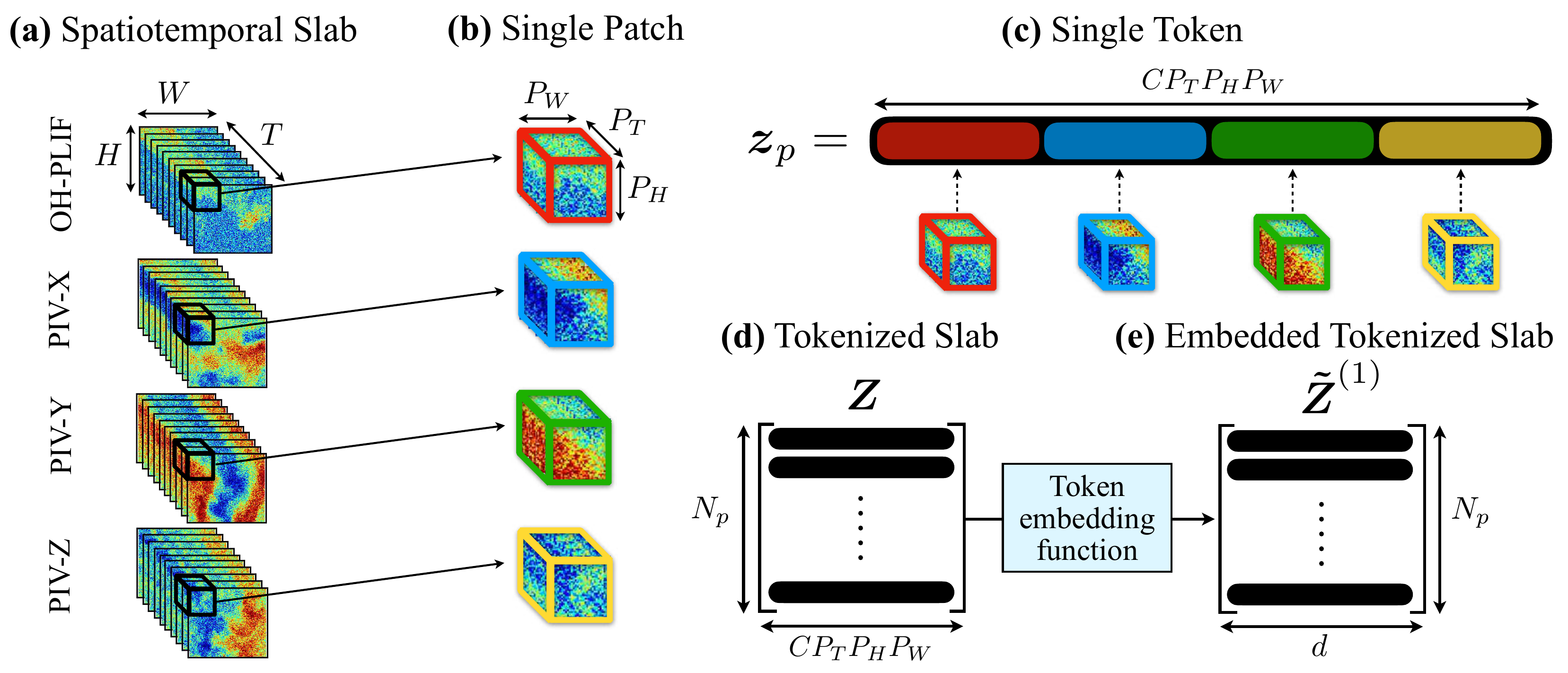}
    \caption{(a) Visualization of a spatiotemporal noised flame slab, showing each feature individually. (b) Visualization of a single spatiotemporal patch from the slab for each of the respective features. (c) Illustration of a single token $\bm{z}_p$ produced as a flattened representation of the patch. (d) Tokenized representation of the slab, which is a matrix comprised of all $N_p$ tokens along the rows and the input token dimensionality along the columns. (e) Embedded tokenized slab, resulting from the token embedding function (Eq.~\ref{eq:tok_embedding}), transforming token dimensionality to the model dimension $d$.}
    \label{fig:Patching}
\end{figure}

\underline{\textbf{Tokenization:}} A spatiotemporal patching process \cite{vivit} is applied to each noised slab to produce a set of tokens, analogous to image-based patching used in vision transformers \cite{vision_transformer}. This is done in two stages. First, the input slab ${\bm z}$ is partitioned into non-overlapping spatiotemporal cuboids (patches) of size $P_T$, $P_H$, and $P_W$ along the time, height, and width axes respectively, producing a patch grid of size $T/P_T \times H/P_H \times W/P_W = N_p$ total patches. Then, each patch is flattened into a vector to produce a token -- when done for all patches, the full tokenized representation of the slab ${\bm z}$ is recovered, and is given by the matrix ${\bm Z} \in \mathbb{R}^{N_p \times C  P_T  P_H  P_W}$. Row $p$ of this matrix contains the $p$-th token ${\bm z}_p \in \mathbb{R}^{C P_T  P_H  P_W}$, which is a flattened representation of the spatiotemporal data within one patch. Figure~\ref{fig:Patching} illustrates this process, which is effectively a discretization procedure for the spatio-temporal slab. Each patch/token coincides with a point in space-time, and the size of the patch (a type of resolution on the space-time grid) sets the minimum length and time-scale that can be resolved by the cross-attention operations in the transformer.

\underline{\textbf{Token and Conditioning Variable Embedding:}} Before entering the transformer blocks, a given input token $\bm{z}_p \in \mathbb{R}^{C  P_T  P_H P_W}$ is first lifted into an \textit{embedded} token representation $\tilde{\bm{z}}_p \in \mathbb{R}^{d}$, where $d$ is the model (or embedding) dimension. This is given by the function
\begin{equation}
    \label{eq:tok_embedding}
    \tilde{\bm{z}}_p = {\cal E}_p({\bm z}_p;\theta_p)
\end{equation}
that operates on each token independently, where ${\theta}_p$ is the learnable parameter set for ${\cal E}_p$. Here, a linear embedding function is used for ${\cal E}_p$ following a bottleneck embedding approach \cite{jit}, wherein a two-layer linear feed-forward network lifts the token into embedding dimension $d$. To preserve awareness of spatiotemporal location within a given slab, the $d$ embedded token values are updated by adding positional encodings using a standard sinusoidal-based approach \cite{vaswani}.

Along with patch embedding, the model relies on two additional conditioning inputs during training: the pseudo-time $\tau$ and the integer class label $y$. To facilitate conditioning, these scalars are also lifted into the model dimension $d$ before entering the transformer block, using separate embedding functions given by
\begin{equation}
\label{eq:tcon}
    \tilde{\bm{\tau}} = {\cal E}_\tau(\tau;\theta_\tau) \in \mathbb{R}^d
\end{equation}
and
\begin{equation}
\label{eq:ycon}
    \tilde{\bm{y}} = {\cal E}_y(y;\theta_y) \in \mathbb{R}^d
\end{equation}
for $\tau$ and $y$, respectively, with parameters $\theta_{\tau}$ and $\theta_{y}$. The embedding function in Eq.~\ref{eq:tcon} is a two-layer feed-forward network (FFN) with a sigmoid linear unit (SiLU) activation applied to a sinusoidal embedding of the input, following diffusion modeling conventions \cite{ddpm,vaswani}. The embedding function in Eq.~\ref{eq:ycon} is a learnable lookup table that maps each discrete class label (0 or 1) to a trainable vector in $\mathbb{R}^d$, resembling token embedding tables used in natural language processing \cite{word2vec}. Once embedded, these vectors are summed into a single \textit{conditioning vector} $\mathbf{c}$,
\begin{equation}
    \label{eq:conditioning_Vec}
    \mathbf{c} = \tilde{\bm{y}} + \tilde{\bm{\tau}} \in \mathbb{R}^{d}.
\end{equation}
As the name implies, the conditioning vector in Eq.~\ref{eq:conditioning_Vec} serves as an auxiliary embedded token that conditions the transformer block (described next) on both pseudo-time location and the class label, and is fixed for the entire slab.

\underline{\textbf{Conditional Transformer Block:}} The conditional transformer used here follows the AdaLN-Zero strategy \cite{dit} -- this style of conditioning was selected due to its demonstrated success in recent computer vision and scientific machine learning applications for diffusion modeling \cite{jit,wang2026fundiff}. At a high level, the conditional transformer block is given by the general operation
\begin{equation}
    \label{eq:transformer}
    \tilde{\bm{Z}}^{(l+1)} = {\cal F}(\tilde{\bm{Z}}^{(l)},\bm{c}; \theta_{{\cal F}}^{(l)}), \quad l=1,\ldots, L,
\end{equation}
where ${\cal F}$ is a transformer, $\tilde{\bm Z}^{(l)} \in \mathbb{R}^{N_p \times d}$ is the embedded tokenized time slab at layer $l$, ${\bm c}$ is the conditioning vector defined in Eq.~\ref{eq:conditioning_Vec}, and $\theta_{{\cal F}}^{(l)}$ represents the learnable parameter set of the transformer at layer $l$. There are a total of $L$ blocks in the architecture -- the action of a transformer block updates the token embeddings from ${\tilde{\bm Z}^{(l)}}$ to  $\tilde{\bm Z}^{(l+1)}$, with the inputs at the first layer ${\tilde{\bm Z}^{(1)}}$ given by the initial embedded set of tokens. This operation is recursively applied $L$ times to produce the post-transformer output $\tilde{\bm Z}^{(L)}$. It is noted that while each of the $L$ transformer blocks is independently parameterized, the same conditioning vector $\bm c$ is used for all blocks, allowing for sustained injection of the necessary conditional information into each of the block's operations.

\begin{figure}
    \centering
    \includegraphics[width=0.4\linewidth]{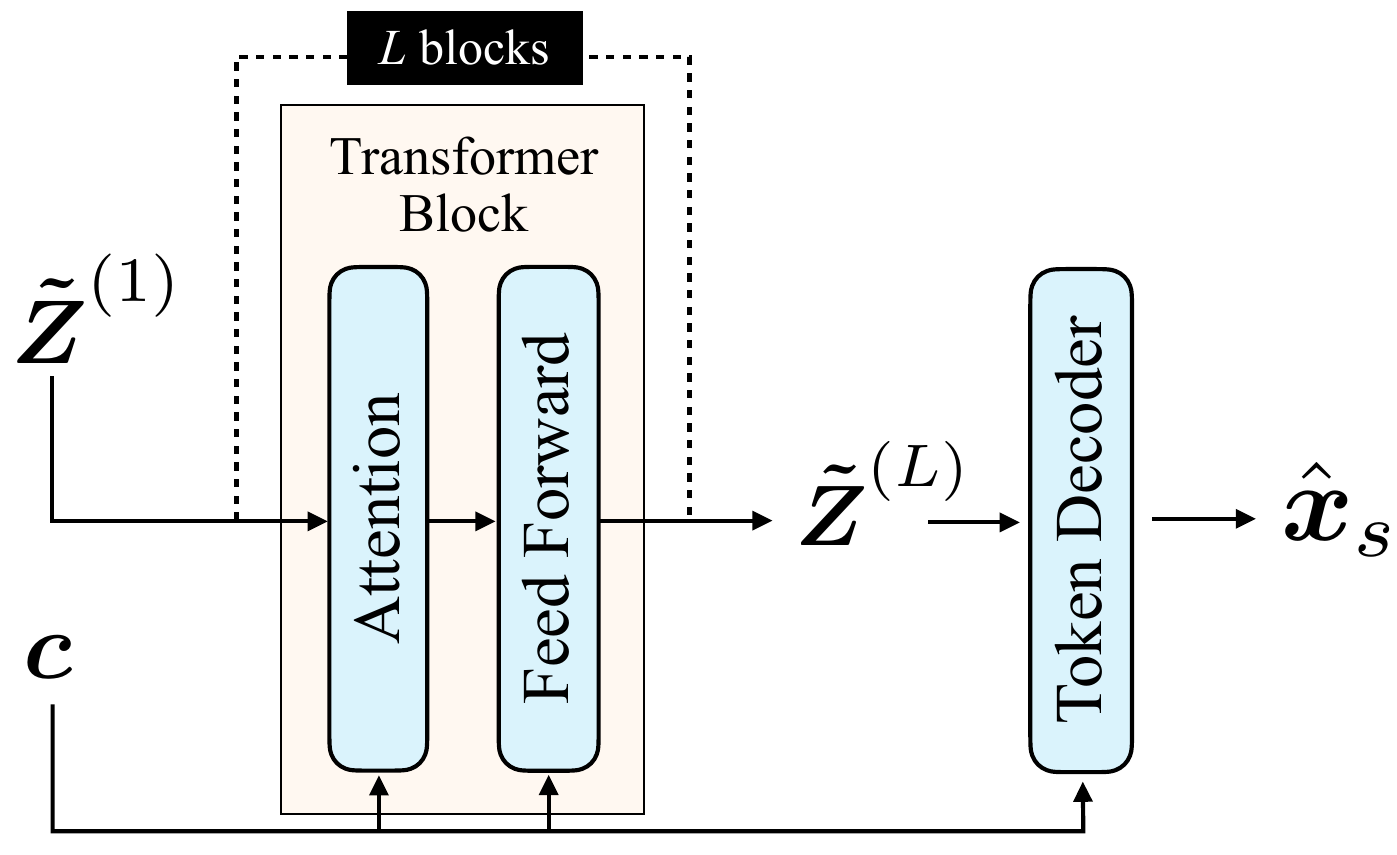}
    \caption{Schematic of the architecture and conditioning strategy following the AdaLN-Zero approach \cite{jit,dit}. The initial embedded tokenized slab $\tilde{\bm Z}^{(1)}$ gets fed into $L$ transformer blocks before passing through the token decoder to produce the final x-prediction (denoised slab), $\bm{\hat x}_s$. The conditioning vector ${\bm c}$ is injected in the transformer block (both attention and feed-forward steps) and the token decoder (see text for details).}
    \label{fig:architecture}
\end{figure}

Inside of each transformer block ${\cal F}$ is an attention step followed by a feed-forward step:
\begin{subequations}
\label{eq:transformer_block}
\begin{align}
    \text{Attention:} \quad & \tilde{\bm{Z}}^{(l)}_{*} = \tilde{\bm{Z}}^{(l)} + \bm{\alpha}(\bm c; \phi^{(l)}_{\text{SDPA}}) \odot \text{SDPA}\big(\hat{\bm{Z}}_{\text{SDPA}}^{(l)}; \theta^{(l)}_{\text{SDPA}}\big), \label{eq:SDPA} \\
    \text{Feed-forward:} \quad & \tilde{\bm{Z}}^{(l+1)} = \tilde{\bm{Z}}^{(l)}_{*} + \bm{\alpha}(\bm c; \phi^{(l)}_{\text{FFN}}) \odot \text{FFN}\big(\hat{\bm{Z}}_{\text{FFN}}^{(l)}; \theta^{(l)}_{\text{FFN}}\big),
    \label{eq:FFN}
\end{align}
\end{subequations}
where
\begin{subequations}
\label{eq:adaLN}
\begin{align}
    \hat{\bm{Z}}_{\text{SDPA}}^{(l)} &= \big(1 + \bm{\gamma} (\bm c; \phi^{(l)}_{\text{SDPA}})\big) \odot \text{LayerNorm}\big(\tilde{\bm{Z}}^{(l)}\big) + \bm{\beta}(\bm c; \phi^{(l)}_{\text{SDPA}}), \label{eq:adaLN_SDPA} \\
    \hat{\bm{Z}}_{\text{FFN}}^{(l)} &= \big(1 + \bm{\gamma}(\bm c; \phi^{(l)}_{\text{FFN}})\big) \odot \text{LayerNorm}\big(\tilde{\bm{Z}}^{(l)}_{*}\big) + \bm{\beta}(\bm c; \phi^{(l)}_{\text{FFN}}). \label{eq:adaLN_FFN}
\end{align}
\end{subequations}
The attention step is given by Eq.~\ref{eq:SDPA}. Here, ``SDPA'' is the multi-head scaled dot-product attention function with parameters $\theta^{(l)}_{\text{SDPA}}$ \cite{vaswani}. The input to SDPA is the normalized token matrix $\hat{\bm{Z}}_{\text{SDPA}}^{(l)}$, whose construction is given by Eq.~\ref{eq:adaLN_SDPA}, which follows the AdaLN procedure described further below. Per Ref.~\cite{jit}, alongside the pre-normalization step in Eq.~\ref{eq:adaLN_SDPA}, query-key normalization \cite{qk_norm} and axial rotary positional encodings (RoPE) \cite{SU2024127063} are applied before entering the attention operation (RoPE is applied axis-wise along the spatial and temporal token grid axes). Ultimately, the attention step acts as a non-local operation in space-time, since the MHSA function models complex information exchange pathways among all tokens at once, and tokens here coincide with spatiotemporal patches -- the attention step can be interpreted as a globally-informed update on the token embeddings, leveraging all space-time information in the slab relevant to the x-prediction-based denoising task for flame dynamics. For additional information on the internal MHSA operations, the reader is directed to Ref.~\cite{vaswani}. 

The output of the attention step is then provided as input to the feed-forward step, which is given by Eq.~\ref{eq:FFN}. Here, ``FFN'' is a transformer-specialized feed-forward network consisting of a gated linear unit with a SiLU activation (SwiGLU FFN \cite{swiglu_ffn}) applied to the normalized token matrix $\hat{\bm{Z}}_{\text{FFN}}^{(l)}$, given by Eq.~\ref{eq:adaLN_FFN}. Following standard transformer conventions, the hidden dimensionality of the FFN in Eq.~\ref{eq:FFN} is set to $4d$, where $d$ is the model dimension. It is noted that unlike the attention step, the FFN in Eq.~\ref{eq:FFN} acts token-wise -- it is a localized operation on each token independently. As such, the transformer block consists of a \textit{global} step (Eq.~\ref{eq:SDPA}) that mixes information contained in all tokens in the slab, followed by a \textit{local} step (Eq.~\ref{eq:FFN}) that mixes information contained in the individual features/dimensions within each token independently \cite{vaswani}.

In Eqs.~\ref{eq:transformer_block}-\ref{eq:adaLN}, conditioning is introduced through the AdaLN-Zero procedure \cite{dit} via three types of functions appearing in both the attention and feed-forward steps: (1) the gating function $\bm\alpha$ applied to the updated token matrices produced by SDPA and FFN; (2) the shift function $\bm\beta$, used in the layer normalization procedure for the SDPA and FFN inputs in Eq.~\ref{eq:adaLN}; and (3) the scale function $\bm\gamma$, used in the same layer normalization procedure. Each of these functions produces a $d$-dimensional output --  $\bm\alpha(\bm c; \cdot), \bm\gamma(\bm c; \cdot), \bm\beta(\bm c; \cdot) \in \mathbb{R}^d$ -- which modulates the $d$ token dimensions in a manner informed by the conditioning vector $\bm c$ (the $\odot$ symbol in Eqs.~\ref{eq:transformer_block}-\ref{eq:adaLN} denotes token dimension-wise multiplication). These functions are independently parameterized single-layer feed-forward networks with SiLU activations, with parameters indicated by $\phi^{(l)}_{\text{SDPA}}$ and $\phi^{(l)}_{\text{FFN}}$ (different parameters used in each of the $L$ transformer blocks, and for each of the attention and feed-forward steps within a block). It is emphasized that these functions take the conditioning vector $\bm c$ as input, allowing both the inputs and outputs of the SDPA and FFN functions to be conditioning vector-aware without modification of the underlying SDPA and FFN functions themselves. The reader is directed to Ref.~\cite{dit} for additional detail. In Eq.~\ref{eq:adaLN}, the ``LayerNorm'' operation follows the root mean square layer normalization approach \cite{rmsnorm}.

\underline{\textbf{Token Decoding:}} Once the tokens have cleared through the $L$ transformer blocks, the result is a new set of token embeddings $\tilde{\bm Z}^{(L)} \in \mathbb{R}^{N_p \times d}$ (the final embeddings). In the token decoding stage, the final embeddings are linearly projected back into the desired output dimensionality -- in other words, tokens in the final embeddings (which live in the model dimension $d$) are projected back to the desired output dimension for the respective spatiotemporal patch (which here is $C P_T P_H P_W$). Denoting a single token from the final embedding matrix as $\tilde{\bm z}^{(L)}_p \in \mathbb{R}^d$, the decoding operation is expressed as
\begin{equation}
\label{eq:decode}
    \bm z_p = {\mathcal{D}}\big(\hat{\bm{z}}^{(L)}_p; \theta_{\cal D}\big) \in \mathbb{R}^{C P_T P_H P_W},
\end{equation}
where $\cal D$ is a learnable linear projection with parameter set (weights) $\theta_{\cal D}$ fixed for all tokens. The input to the decoder is the normalized final embedded token,
\begin{equation}
\label{eq:decode_adaLN}
    \hat{\bm z}^{(L)}_p = \big(1 + \bm\gamma(\bm c; \phi_{\cal D})\big) \odot \text{LayerNorm}\big(\tilde{\bm z}^{(L)}_p\big) + \bm\beta(\bm c; \phi_{\cal D}),
\end{equation}
following the same AdaLN-based conditioning pattern used in the transformer block. That is, a conditioning-dependent normalization step is applied to $\tilde{\bm z}^{(L)}_p$ before application of the decoder function $\cal D$, using the same shifting and scaling strategy as in Eq.~\ref{eq:adaLN}, but with different parameters in the respective shift/scale functions. As a final step, once all tokens are decoded, the decoded slab is reshaped to the original 4-D spatiotemporal layout, recovering the x-prediction $\hat{\bm{x}}_s$.

\section{Results}
\label{sec:results}

With the diffusion modeling approach described in Sec.~\ref{sec:methods}, analysis and discussion of the model-synthesized spatiotemporal flame data is now presented. The results are separated into two components. First,  Sec.~\ref{subsec:val} centers on analysis of generated data in attached and lifted flame regimes, which serves as a qualitative and quantitative assessment of the model's conditional generation ability. Then, Sec.~\ref{subsec:transition_synthesis} provides assessment and demonstration of the model in an extrapolative context, namely through a novel approach for generating transitioning flame dynamics between attached and lifted states, constituting generation of combustion extreme event data (liftoff and reattachment) that is unseen during training.

Specifically, under a fixed architecture, conditional diffusion models trained to generate two different temporal slab sizes are compared -- $T=10$ (10 frames, termed the $T_{10}$ model) and $T=100$ (100 frames, termed the $T_{100}$ model). Under the same diffusion model framework, different $T$ values requires training different transformers for the x-prediction task (i.e., in the inference stage, the $T_{10}$ model generates $10$ frames of OH-PLIF / PIV fields, whereas the $T_{100}$ model generates $100$ frames). Expansion of the temporal slab size to $T=100$ allows the model to span the necessary dynamical range required for lifted and transitioning flames, and the comparison between $T_{10}$ and $T_{100}$ facilitates study on the impact of increased time-span on the quality of generated dynamics.

Model details are provided in Table~\ref{table:models}. To isolate the effect of slab size on generation quality, the transformer architecture (including the patch size) is fixed for both models: all reported results use the number of transformer blocks $L=8$, and the model embedding dimension $d=1080$. Models are implemented in PyTorch, and both models are trained to 1500 epochs using distributed data parallel (DDP)-based training (per-GPU batch size is 16 for $T_{10}$ and 4 for $T_{100}$) on the Aurora high-performance computing cluster at the Argonne Leadership Computing Facility \cite{aurora}. Each node on Aurora houses 6 Intel Data Center GPU Max Series (Ponte Vecchio) GPUs. On Aurora, a single GPU is split into 2 GPU ``tiles,'' such that each node registers 12 effective GPUs per node via the XPU backend in PyTorch. Two nodes (24 effective GPUs) were used to train the $T_{10}$ model, and ten nodes (120 effective GPUs) were used to train the $T_{100}$ model. Flash attention \cite{flash_attention} was used for SDPA, which eliminates memory constraints associated with the relatively large number of tokens per slab in the $T_{100}$ model at the patch size considered. 

Training leverages learning rate warmup for the first 100 training iterations, with epoch-based cosine learning rate decay scheduling starting at a learning rate of $10^{-3}$ and ending at a learning rate of $10^{-6}$. The AdamW optimizer \cite{adamw} is used with gradient clipping \cite{gradient_clipping}, with clipping threshold set to $1$. Before training, all data is standardized independently for each feature using training data statistics.

\begin{table}[htbp]
\centering
\caption{Settings for the $T_{10}$ and $T_{100}$ models. Both models leverage $L=8$ transformer blocks with model dimension $d=1080$ (resulting in roughly 188M parameters).}
\label{table:models}
\begin{tabular}{lccccc}
\hline
Model & \makecell{Slab Size \\ $(T,H,W)$} & \makecell{Patch Size \\ $(P_T, P_H, P_W)$} & \makecell{Tokens-per-Slab\\ $N_p$} & \makecell{Training Resources \\ (GPUs)} & \makecell{Training Time \\ (Seconds-per-Epoch)} \\ 
\hline
$T_{100}$  & $(100, 128,128)$ & $(2,8,8)$ & 12800 & 120 & 146\\
$T_{10}$ & $(10,128,128)$ & $(2,8,8)$ & 1280 & 24 & 62\\
\hline
\end{tabular}
\end{table}

\subsection{Conditional Generation of Flame Evolution}
\label{subsec:val}

The results below demonstrate the conditional generation capability of $T_{10}$ and $T_{100}$ models in attached and detached flame regimes. First, comparisons of synthetic to real-data counterparts in physical space are analyzed in Sec.~\ref{subsec:RvG}, followed by probe-based temporal analysis in Sec.~\ref{subsec:probes} and statistical comparisons facilitated by proper orthogonal decomposition (POD) in Sec.~\ref{subsec:statistical}.

\subsubsection{Physical Space Visualizations}
\label{subsec:RvG}

\begin{figure}[t]
    \centering
    \includegraphics[width=\linewidth]{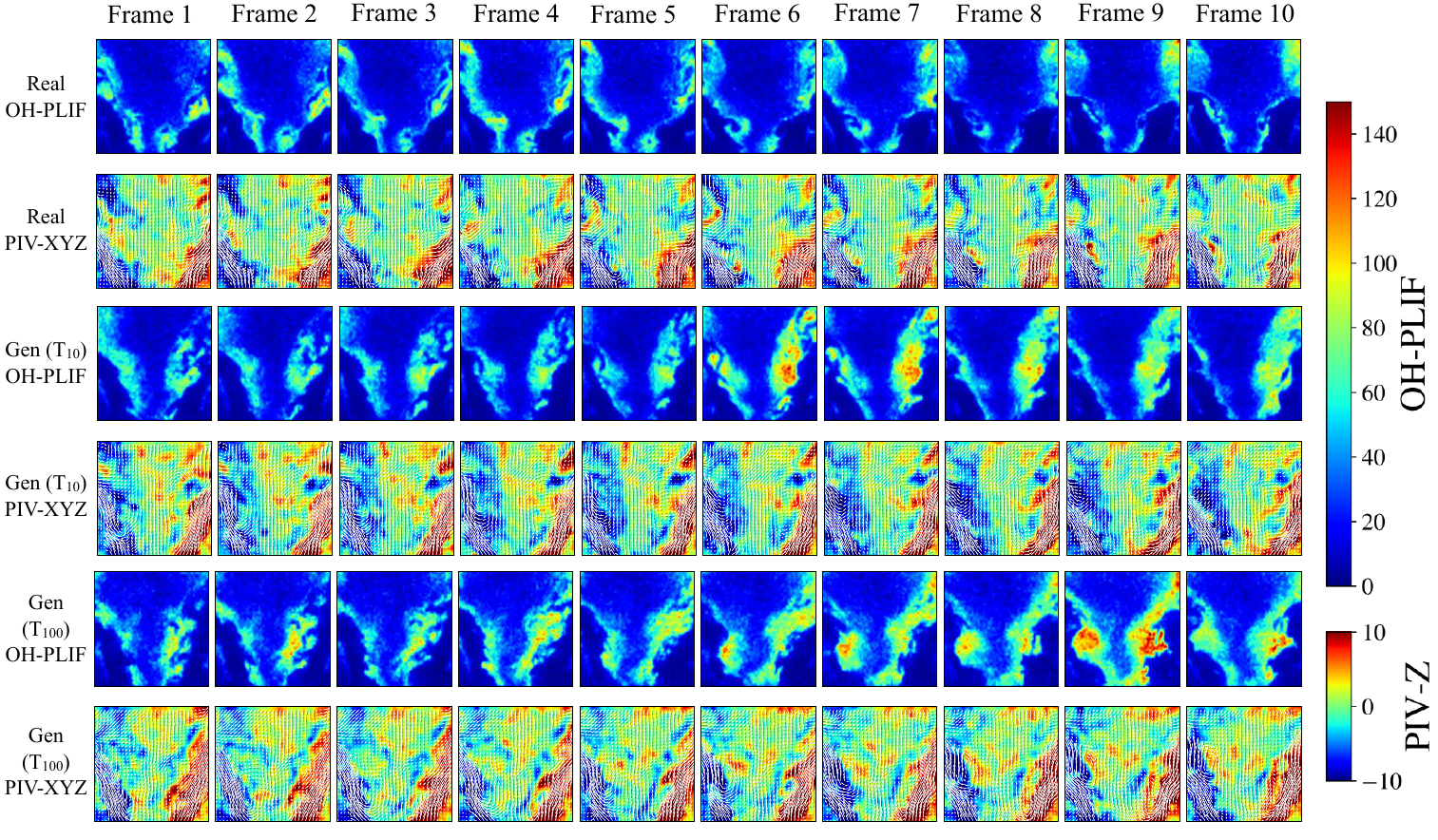}
    \caption{Attached flame trajectory, showing 10 time-ordered frames sourced from a real sequence (first two rows), the $T_{10}$ generative model (next two rows), and the $T_{100}$ generative model (last two rows). Each row pair displays simultaneous OH-PLIF and PIV-Z fields, with velocity vector overlays sourced from PIV-X/Y components.}
    \label{fig:RVG-A}
\end{figure}

Trajectory comparisons for flames in the attached configuration are shown in Fig.~\ref{fig:RVG-A}, comparing 10 sequential frames from a real trajectory to generated counterparts. Shown are simultaneous OH-PLIF and PIV fields, with PIV fields consolidated into single images using PIV-Z visualizations overlaid with velocity vector arrows sourced from PIV-X and PIV-Y data for convenience. While PIV-X and PIV-Y fields are not shown directly, generation quality is overall equivalent to that observed for the PIV-Z fields.

From a qualitative standpoint, Fig.~\ref{fig:RVG-A} illustrates how synthetic flames preserve the large-scale instantaneous features of attached flames. Through visualization of the OH-PLIF fields, both $T_{10}$ and $T_{100}$ models successfully produce flames that are (a) anchored to the burner exit, and (b) characterized by symmetric V-shaped streaks. Inspection of the velocity fields tells a similar story, with synthetic flames capturing the expected anti-symmetric structure in PIV-Z fields alongside the presence of recirculation zones near the strongly positive and negative regions of out-of-plane velocity. Alongside large-scale structure, Fig.~\ref{fig:RVG-A} indicates preservation of complex small-scale features by the diffusion model -- for example, in Frame 1, both $T_{10}$ and $T_{100}$ models generate V-shaped flames with spatially intermittent pockets of OH and velocity field variation about the large-scale V-shaped structure, characteristic of turbulent effects in this configuration. 

It is emphasized that the $T_{100}$ and $T_{10}$ flame trajectories are not expected to perfectly match real counterparts (and one another) due to differences in initial frames, the randomness in generative sampling, and the turbulent nature of the reacting flowfield. In Fig.~\ref{fig:RVG-A}, this leads to apparent visual deviations in certain frames with respect to the real sequence. For example, at Frames 7-10, the $T_{10}$ model shows a striking imbalance in the OH field, with notably higher OH concentration on the right side of the domain relative to the left, a quality not present in the depicted real and $T_{100}$ counterparts. Similarly, at Frames 5-10, the $T_{100}$ model generates flames with higher levels of OH concentration immediately at the burner exit, a quality not present in the depicted real or $T_{10}$ counterparts. While not shown here, these deviations are \textit{not} anomalous artifacts -- similar features are indeed observed in real attached flames, in different trajectories.

\begin{figure}[t]
    \centering
    \includegraphics[width=\linewidth]{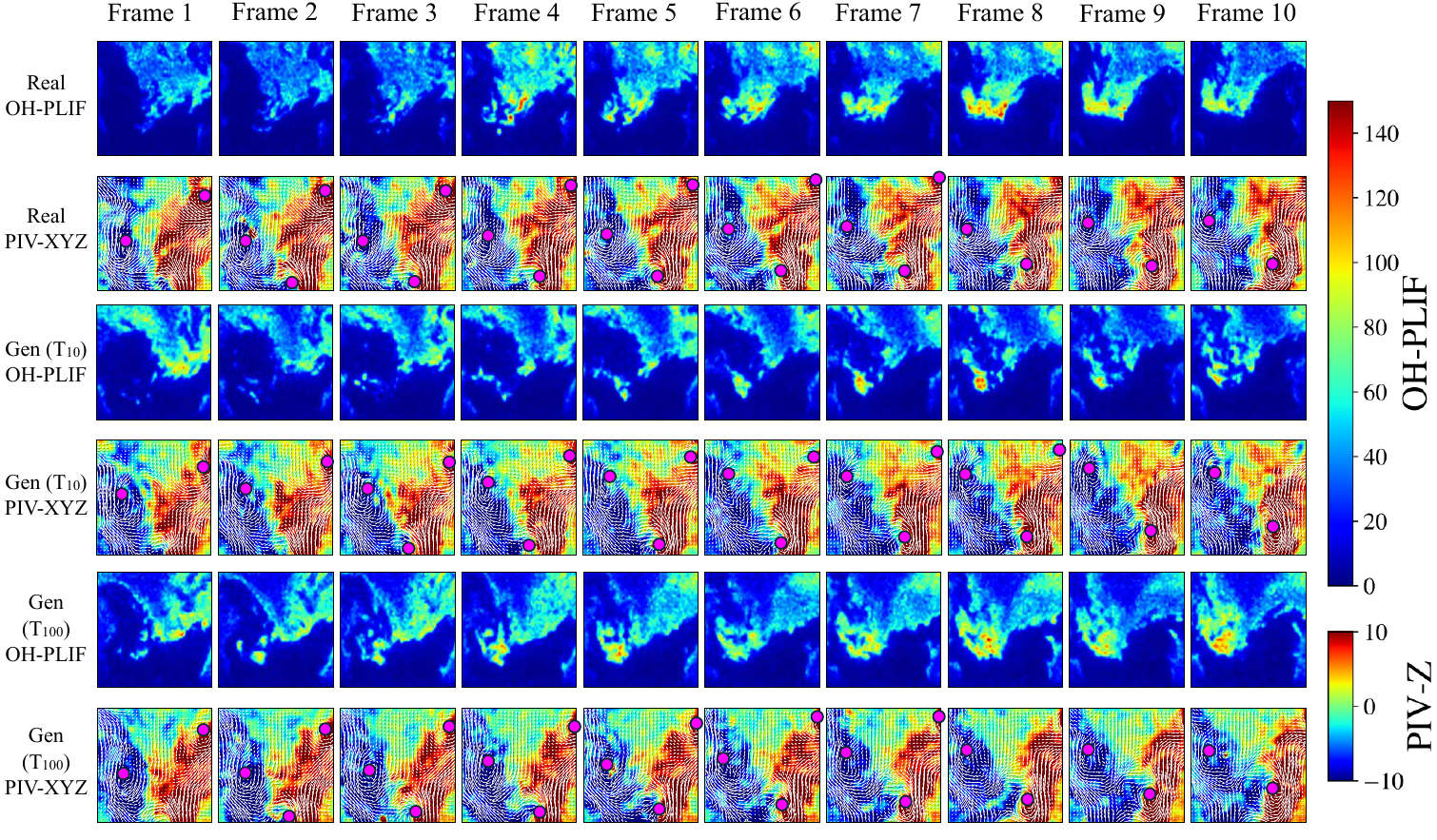}
    \caption{Detached flame trajectory, showing 10 time-ordered frames sourced from a real sequence (first two rows), the $T_{10}$ generative model (next two rows), and the $T_{100}$ generative model (last two rows). Each row pair displays simultaneous OH-PLIF and PIV-Z fields, with velocity vector overlays sourced from PIV-X/Y components. Purple markers on PIV-Z fields indicate recirculation zones.}
    \label{fig:RVG-D}
\end{figure}

Analogous 10-frame trajectories for \textit{detached} flames are shown in Fig.~\ref{fig:RVG-D}. Overall trends are consistent with attached flame counterparts -- both generative models capture the key hook-shaped characteristic of the detached flame, with the expected lack of OH concentration near the burner exit and clear anti-symmetry in out-of-plane velocity streaks with qualitatively similar finer-scale fluctuations present across all frames.

The ability of the generative model to synthesize consistent \textit{dynamics} is more discernible in the detached regime: the generated sequences in Fig.~\ref{fig:RVG-D} depict similar movement of the flame anchor point (seen in OH fields), coupled with the movement of the zig-zag pattern of recirculation zones (seen in the purple markers in the velocity fields) caused by the precessing vortex core (PVC) \cite{AN2019267}. Tracking of these recirculation zones across the 10-frame sequences shows how individual vortices can be seen to evolve and progress vertically in a manner consistent with real detached flames. Constant vertical movement of these vortices creates space for new vortices to develop near the burner exit -- a key detached flame dynamical feature that is also captured in the generated flame trajectory (seen by the markers entering the frame from the base, notably in Frame 2, of the Real and $T_{100}$ cases in Fig.~\ref{fig:RVG-D}).

Ultimately, Figs.~\ref{fig:RVG-A} and \ref{fig:RVG-D} confirm that (a) one can generate qualitatively consistent synthetic simultaneous measurement-based \textit{trajectories} of attached and detached flames using a conditional diffusion model, and (b) the dynamical aspect of generative capability (frame-to-frame change) is modeled to good effect with the spatiotemporal transformer formulation combined with x-prediction. Interestingly, across both Figs.~\ref{fig:RVG-A} and \ref{fig:RVG-D}, both $T_{10}$ and $T_{100}$ models preserve the real flame features to similar levels of accuracy on visual inspection, indicating that the transformer capacity is able to accommodate much of the increase in dynamical complexity when jumping from $T=10$ to $T=100$ frames of generation, which increases the number of tokens processed by the attention layers by an order of magnitude at a fixed patch size (Table~\ref{table:models}).

\begin{figure}
    \centering

    \includegraphics[width=0.4\textwidth, trim={0 0 0 0}, clip]{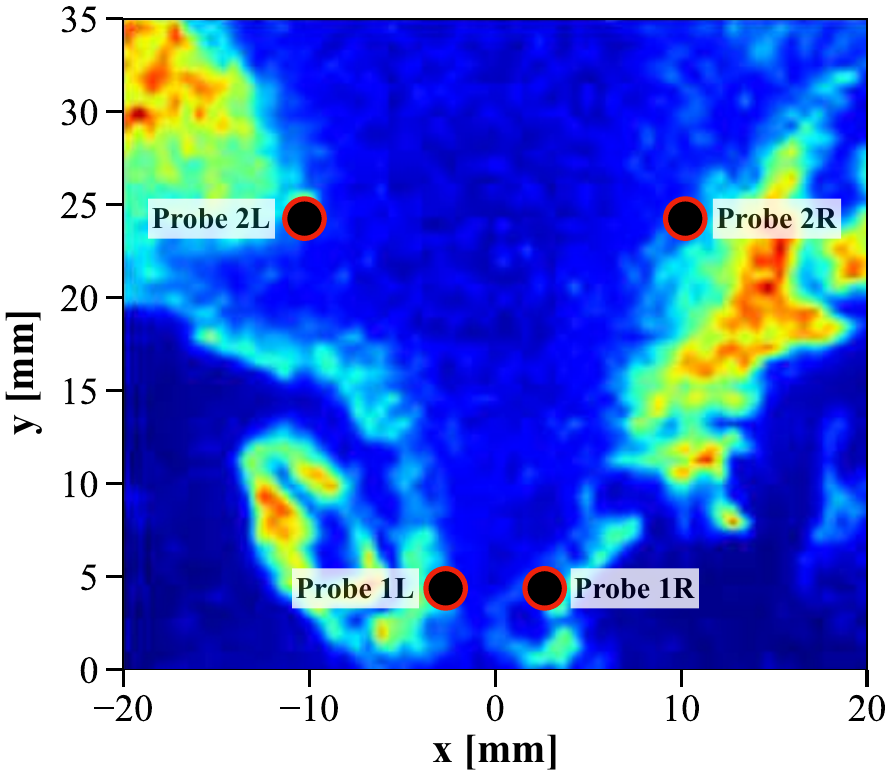}

    \caption{Spatial locations of point probes utilized for the temporal comparisons in Fig.~\ref{fig:noodle_plots}. Probe 1 pair is positioned near the burner exit, whereas Probe 2 pair is located further downstream. Coordinates for each probe are as follows: 1R: $(2.7, 4.3)$, 1L: $(-2.7, 4.3)$, 2R: $(10.2, 24.2)$ and 2L: $(-10.2, 24.2)$ mm. } 
    \label{fig:noodle_probe}
\end{figure}

\begin{figure}[t]
    \centering
    
    \makebox[\textwidth][c]{

            \centering
            \includegraphics[width=\linewidth, trim={0 0 0 0}, clip]{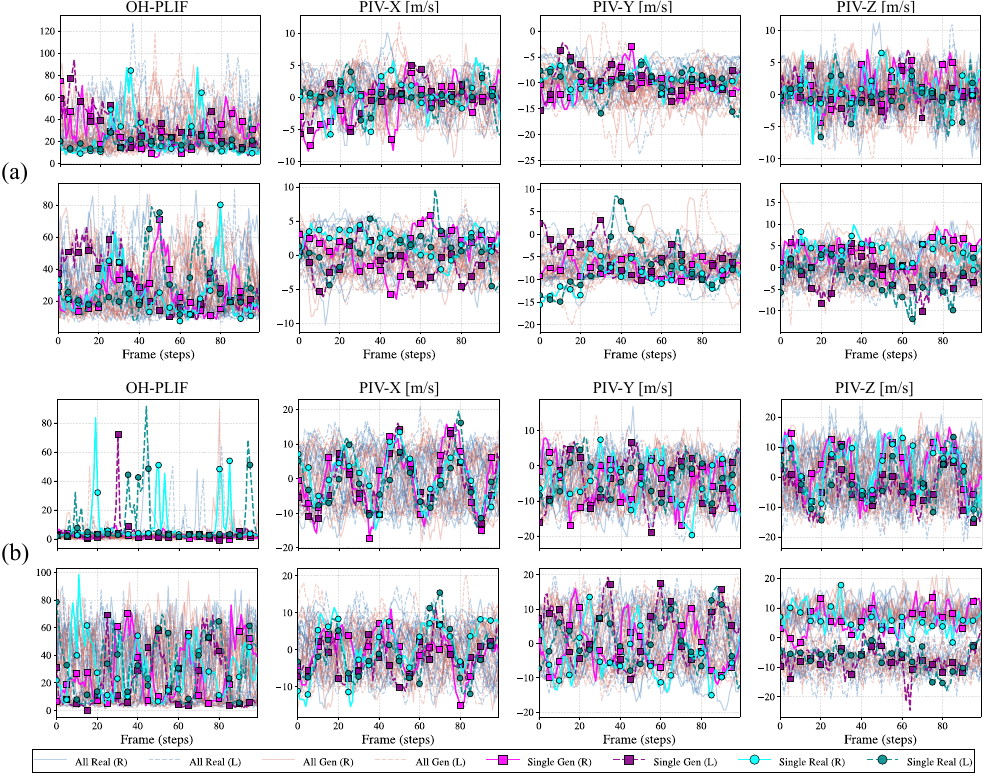}
 }

    \caption{Multi-feature probe time traces comparing real and generated slabs for (a) attached and (b) detached flame regimes. In each sub-figure, the top row corresponds to Probe 1 and the bottom row to Probe 2 (refer to Fig.~\ref{fig:noodle_probe}). Left (L) and right (R) probe signals are represented by dashed and solid lines respectively. Single isolated trajectories are highlighted with bold+symbol curves per the legend, overlaid against all sampled trajectories in corresponding opaque curves. Generated curves come from $T_{100}$ model.}

    \label{fig:noodle_plots}
\end{figure}

\subsubsection{Probe-based Assessment of Synthetic Time Evolution}
\label{subsec:probes}

Given the physical space visualizations described above, a more detailed look into conditional generation capability from a temporal aspect is warranted. To this end, point probes are placed at fixed spatial locations in the spatial ($H\times W$) plane, facilitating comparison of time-evolution of all simultaneous features (OH, PIV-X/Y/Z) between real and generated slabs. This is done here using the $T_{100}$ model, due to its greater temporal span. Probe locations are shown in Fig.~\ref{fig:noodle_probe}. The intent behind the probe placement is to (a) ensure that at least one of the sensor pairs is activated during each state, and (b) allow for the characteristic symmetric and anti-symmetric trends in the flow to be visualized along the temporal axis. Four total probes were deemed sufficient to this end: one symmetric pair near the burner exit (Probes 1L and 1R for left and right parts of the domain, respectively), and another symmetric pair further downstream (Probes 2L and 2R).

Time-evolution plots for the probes are shown in Fig.~\ref{fig:noodle_plots} for all four features, with attached flame evolution in Fig.~\ref{fig:noodle_plots}(a), and detached in Fig.~\ref{fig:noodle_plots}(b). To showcase the dynamical variability in real and generated flame slabs, curves in the plots are sourced from 10 randomly sampled real and generated trajectories, with a single arbitrarily-selected trajectory highlighted for visualization purposes.

For the attached flame state in Fig.~\ref{fig:noodle_plots}(a), symmetric probe pairs produce largely similar signal traces, consistent with the lateral symmetry expected of the flame shape. The exception is in PIV-Z, which shows the key anti-symmetry feature that is captured by the generated slabs: outside of fluctuations, Probe 1R and 1L are roughly equal in magnitude and opposite in sign, mimicking the real signals.  Overall, comparing real and generated samples at the same probe locations shows that the OH-PLIF intensity ranges and temporal fluctuation patterns are similar to real counterparts in attached flames, and the same observation can be made regarding the velocity magnitudes and fluctuation patterns obtained from the PIV channels. Notably, the range in OH-PLIF near the burner exit (top-left plot in Fig.~\ref{fig:noodle_plots}(a)) is properly captured at both upper and lower ends - generated OH signal values observe the same floor as real values at this location, which is well above 0 (as expected for attached flames), and spikes at higher ends (i.e., greater than 100 units of signal intensity) are also captured. 

For the detached flame case shown in Fig.~\ref{fig:noodle_plots}(b), directly apparent is that the PIV signals obtained from both the experimental ground truth and synthesized samples exhibit periodic oscillations (apparent at Probe 1, PIV X/Y/Z), a signal for the swirling vortical structures of the detached regime. Furthermore, analysis of the PIV-Z channel at Probe 2 in the detached regime reveals a more pronounced anti-symmetric distribution that is captured by the synthetic data: both the real and generated probe pairs display the expected highly negative value distribution on the left side and a positive value distribution on the right side, aligning with the visual counterparts in Fig.~\ref{fig:RVG-D}. Additionally, the juxtaposition of Figs.~\ref{fig:noodle_plots}(a) (attached) and Figs.~\ref{fig:noodle_plots}(b) (detached) further emphasizes the conditional generation capability of the model. For example, the generated OH signal distribution at Probe 1 in the detached flames recovers the dominant mode very close to 0 signal intensity; at Probe 2, the generated distribution also recovers the much wider range in OH distribution in the detached regime versus the attached.

Overall, the dynamical trends in Fig.~\ref{fig:noodle_plots} present in the real data are captured by the generated data. Coupled with the findings in Sec.~\ref{subsec:RvG}, this highlights the ability of the diffusion modeling framework to synthesize new flames with spatiotemporally-consistent flow patterns in different and highly distinct flame evolution regimes (attached and detached).

\subsubsection{Statistical Evaluation of Synthetic Flames}
\label{subsec:statistical}

To augment the above analysis, the quality of synthetic flame data is further examined using statistical analysis via proper orthogonal decomposition (POD). Briefly, POD is a modal decomposition technique that optimally decomposes a set of flow snapshots into frozen spatial modes (the POD modes) and time-evolving coefficients (the POD coefficients, one for each mode) \cite{taira_pod}. The collection of POD modes, obtained from eigenvectors of the snapshot covariance matrix, forms an orthogonal basis, with corresponding eigenvalues supplying the energy (variance preserved) along the direction of the mode. Ultimately, each mode constitutes a ``flow direction'', and projection of an instantaneous snapshot onto the mode provides the POD coefficient at the same time instant. For the $k$-th mode, at some time instant $t$, this is given by $a_k(t) = \langle \bm {x}(t), {\bm \phi}_k \rangle$, where $a_k(t)$ is the scalar coefficient, ${\bm x}(t)$ is the instantaneous snapshot, and ${\bm \phi}_k$ is the POD mode. The coefficient measures the degree of alignment of the instantaneous snapshot with the flow features present in the corresponding mode, in a spatially global sense. Additional detail on POD is out of scope here; for full details and algorithms, the reader is directed to Ref.~\cite{taira_pod}.

\begin{figure}
    \centering
    \includegraphics[width=0.8\textwidth]{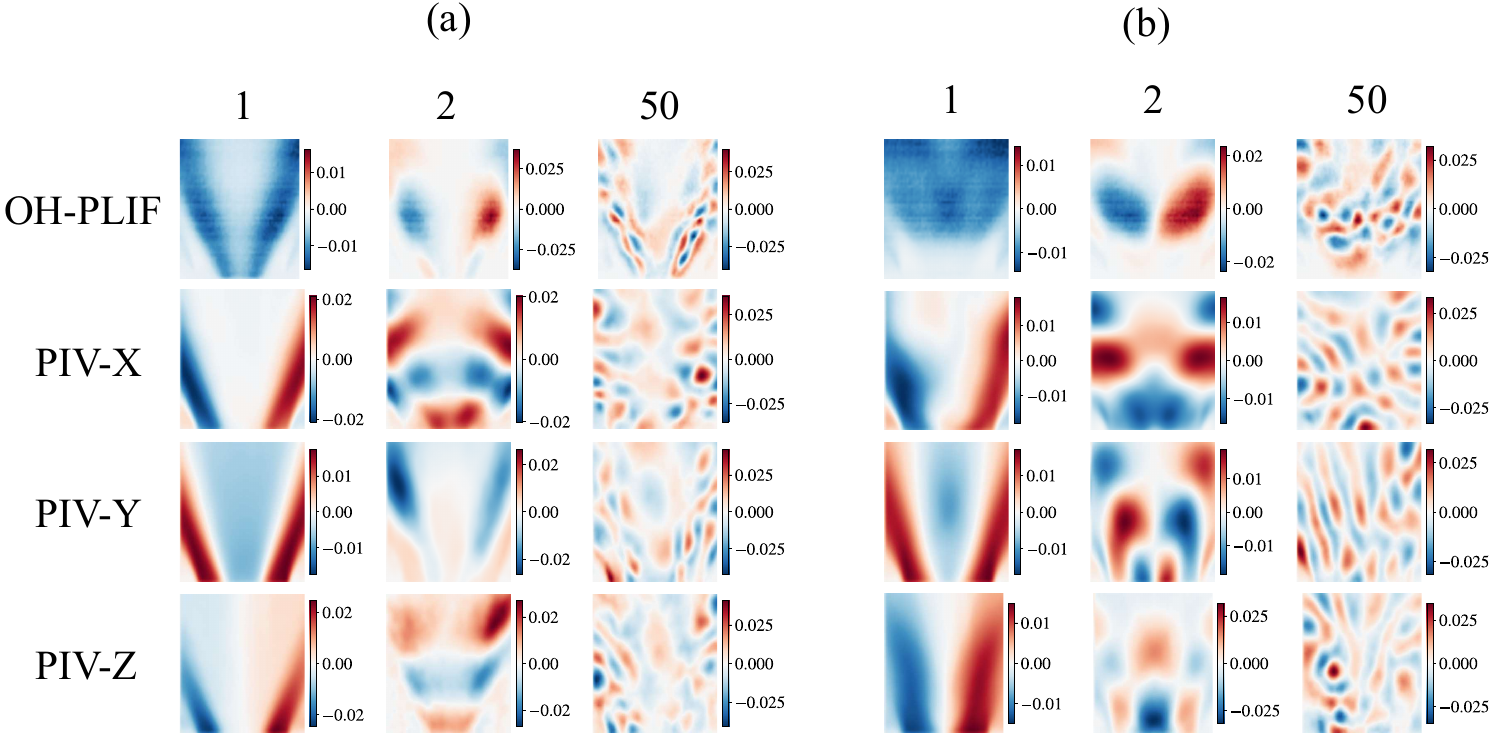}
    \caption{POD modes 1, 2, and 50
    for the (a) attached, and (b) detached flame regimes, produced using real data. Rows indicate feature, and columns indicate mode ID. Modes are ordered in descending order of energy content.}
    \label{fig:pod_modes}
\end{figure}

Here, in a first step, POD is executed independently for the attached and detached flame classes, and for each feature, \textit{using real data} (with means included). This results in a fixed set of POD modes, a subset of which is shown in Fig.~\ref{fig:pod_modes}. Then, in a second step, the real POD modes are used to produce POD coefficients for \textit{both real and generated} data by projecting the data onto the real modes (i.e., the real modes provide a fixed reference frame that enables statistical comparisons through the coefficients, which change with the data being projected).

\begin{figure}
    \centering
    \centering
    \includegraphics[width=1\linewidth, trim={0 0 0 0}, clip]{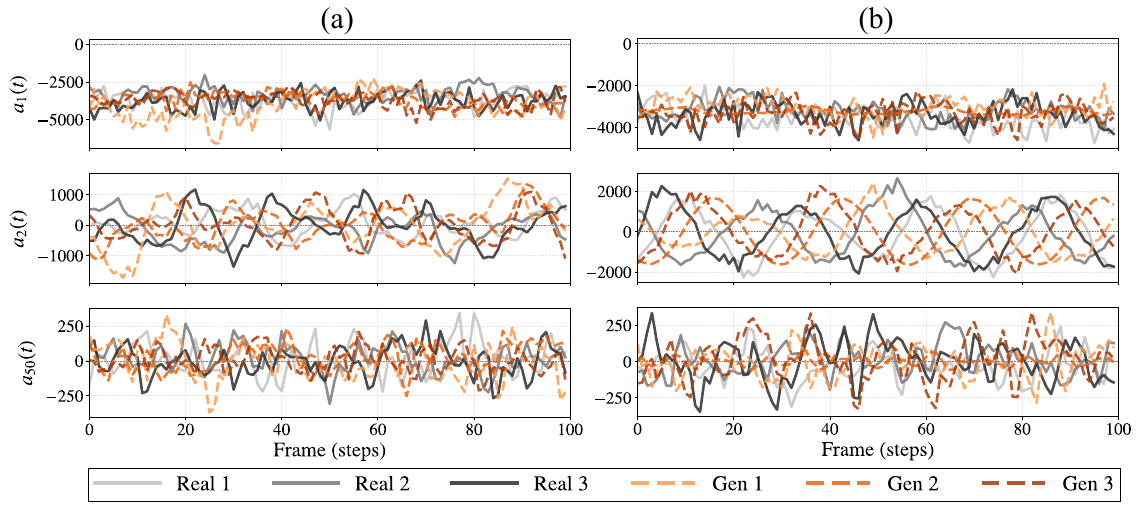}

    \caption{Evolution of POD Time Coefficients of OH-PLIF signal with time for $T_{100}$ dataset. Three random real and generated slabs are shown for (a) attached and (b) detached flame regimes. Within a given subfigure, each subplot corresponds to a specific mode and shows the time series of the corresponding POD coefficient for both real and generated data.}
    \label{fig:pod_time_coefficients}

\end{figure}

Evolution of real and generated OH-PLIF POD coefficients for the three modes (Modes 1, 2, and 50, see Fig.~\ref{fig:pod_modes}) is shown in Fig.~\ref{fig:pod_time_coefficients}. Comparison of the real and generated POD coefficient trajectories provides a global assessment of the captured dynamics at varying length-scales contained in the modes: $a_1(t)$ captures the mean alignment (largest scales), $a_2(t)$ represents large-scale patterns outside of the mean, and $a_{50}(t)$ represents alignment with smaller-scale coherent structures. Trends in Fig.~\ref{fig:pod_time_coefficients} show that time evolution at these multiple scales in the generated slabs closely resemble real counterparts -- similar shifts in coefficient ranges and oscillation patterns are observed in time. In particular, there is consistent multi-scale variation in generated samples in both attached and detached regimes: the diffusion model balances the preservation of the coefficient temporal structure while enforcing sample diversity at multiple scales.

\begin{figure}[htbp] 
  \centering
    \includegraphics[width=1\linewidth, trim={0 0 0 0}, clip]{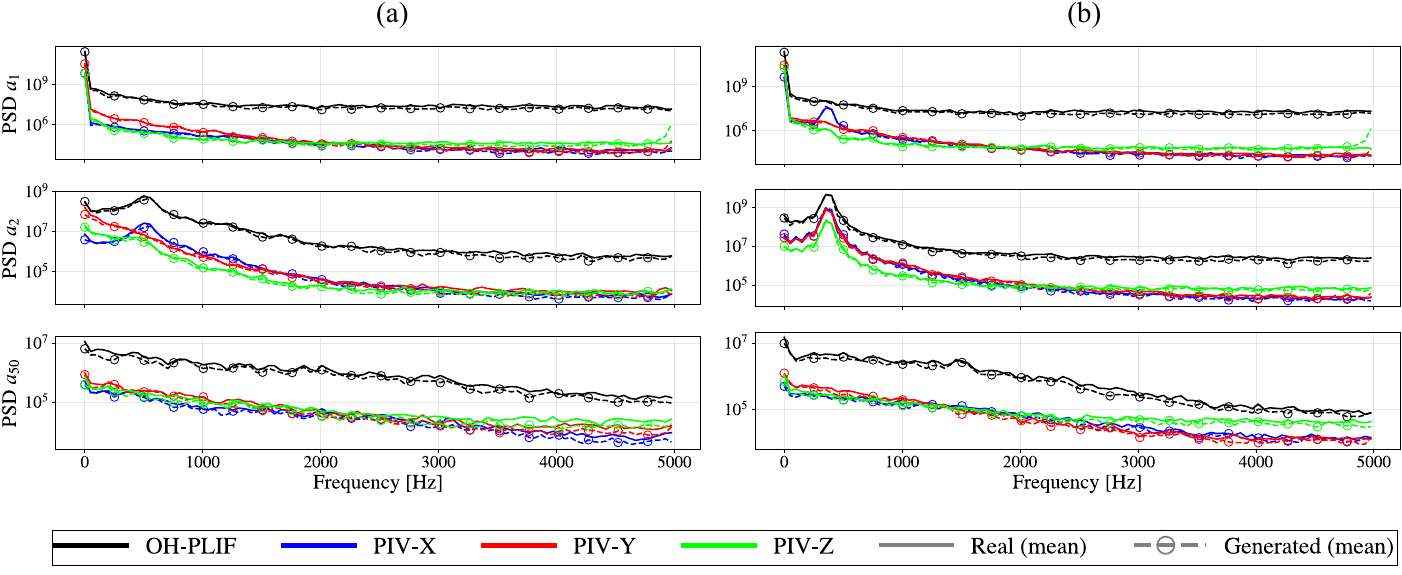}
    \caption{Mean PSD of POD temporal coefficients for selected modes of the $T_{100}$ dataset (averaged over an ensemble of 300 real and 300 generated slabs), for (a) attached and (b) detached flame regimes. Each panel shows the PSD of the temporal coefficients for modes 1, 2, and 50 (shown in Fig.~\ref{fig:pod_modes}).}
    \label{fig:pod_psd}
\end{figure}

Mean power spectral density (PSD) curves for the same three POD coefficients in both attached and lifted regimes, for all four components, are shown in Fig.~\ref{fig:pod_psd}, which provides a global representation of temporal consistency at multiple length and timescales. Overall, the $T_{100}$-generated PSD curves are largely consistent with real counterparts in both attached (Fig.~\ref{fig:pod_psd}(a)) and detached (Fig.~\ref{fig:pod_psd}(b)) regimes for all four components. In particular, key large-scale temporal structure is captured -- evidenced by the peak in frequency near $400$~Hz in the detached regime for the second POD coefficient, which corresponds to the PVC structure \cite{AN2019267} (the periodic oscillations in the corresponding POD coefficients are evident in Fig.~\ref{fig:pod_time_coefficients}(b)).

While there is no drastic deviation, the PSD curves reveal slight undershoots in the generated data, with the most prominent undershoots occurring at coefficient 50 (smaller scales). When considering the fact that larger-scale spatiotemporal features are almost perfectly captured (e.g., the frequency peaks in the detached flame in the first and second coefficient), such finer-scale deviations imply a difficulty in the denoising mechanism at the finest scales. While this property requires further investigation, it can be attributed to higher signal-to-noise ratios of the larger-scale spatiotemporal features, as well as the x-prediction based method itself (the spatiotemporal transformer), whose cross-attention fidelity is constrained by the patch size (the patch size here spans a cuboid of 2 frames in the temporal direction and 8 pixels in the spatial direction). Reductions in patch size have been shown in computer vision applications to improve image generation quality \cite{dit} -- while expensive, such modifications may help mitigate this deviation.

\begin{figure}[t]
    \centering
    
    \makebox[\textwidth][c]{\includegraphics[width=1\textwidth, trim={0 0 0 0}, clip]{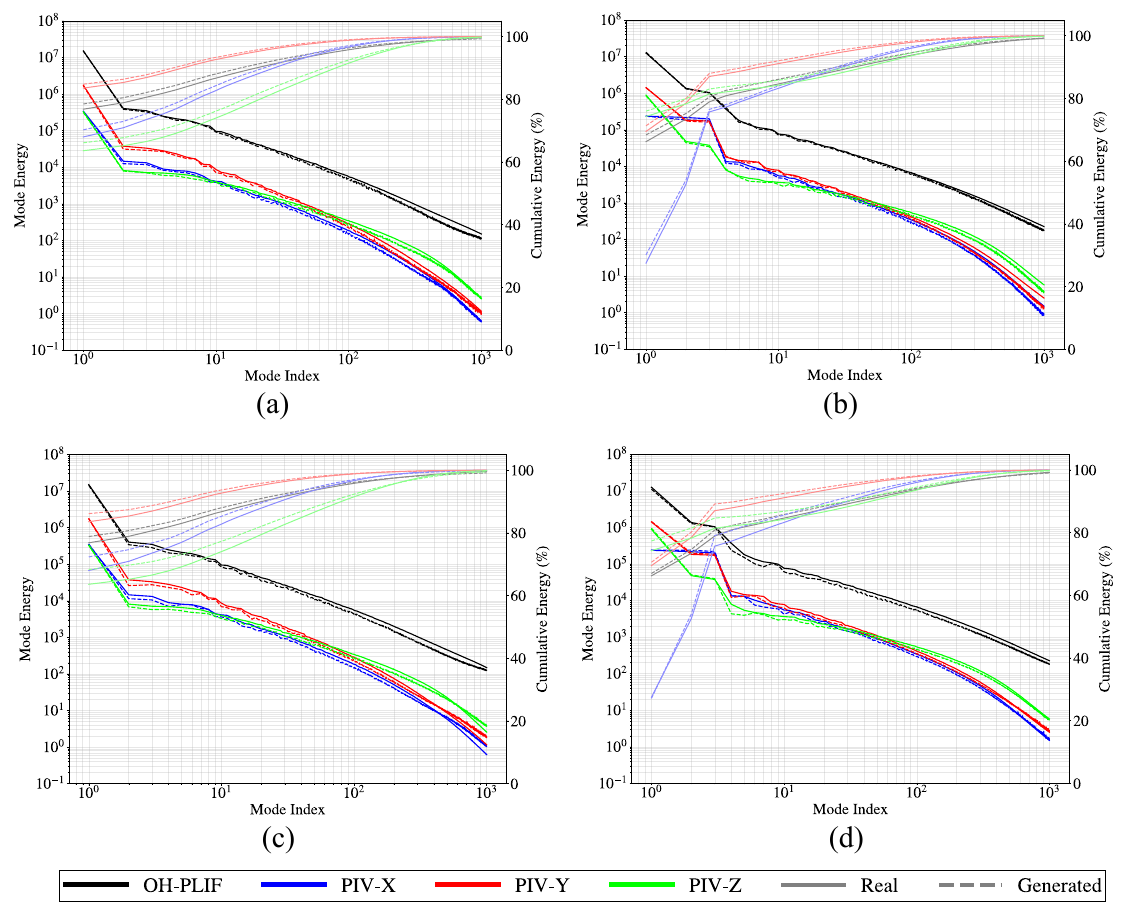}}

    \caption{Real and synthetic POD spectra for the OH-PLIF and PIV-X/Y/Z modalities, computed using a fixed POD basis produced from the real data. The primary $y$-axis shows the per-mode energy content, and the secondary $y$-axis shows the cumulative energy (\%) as a function of mode index. In each plot, the real spectra are given by solid lines, and synthetic spectra are given by dashed lines: (a) $T_{10}$ model, attached regime; (b) $T_{10}$ model, detached regime; (c) $T_{100}$ model, attached regime; (d) $T_{100}$ model, detached regime.}
    \label{fig:pod_spectra}
\end{figure}

Alongside temporal statistical characteristics of generated data samples, assessment of spatial statistical consistency is required to gain a more complete understanding of generation capability. To this end, mean POD spectra are shown in Fig.~\ref{fig:pod_spectra} for both $T_{10}$ and $T_{100}$ models in attached and detached regimes; the POD spectrum is a representation of the \textit{spatial} variation of pixel values in each of the features along the respective modes. Using the fixed POD basis produced from the real data as described above, each plot in Fig.~\ref{fig:pod_spectra} displays individual mode energy (which comes from the square of the POD coefficient) and cumulative mode energy as a function of mode index for all four features (OH-PLIF and PIV-$X$, $Y$, $Z$).

The POD spectra show that both $T_{10}$ and $T_{100}$ models capture a very similar statistical distribution of instantaneous flames, which can be attributed to the fixed patch size used to train them. However, on closer inspection, at the lower mode indices (larger spatial scales), the $T_{10}$ model aligns closer to the real distribution when compared with the $T_{100}$ model (this deviation is more apparent in the cumulative energy profiles) in both attached and lifted regimes, implying that the increased frame count in the generation task for $T_{100}$ (greater temporal generation fidelity) comes at a slight penalty of spatial reconstruction accuracy. Interestingly, increasing generation capacity from 10 to 100 frames results in a consistent increase (vertical shift) in the energy in synthetic flames \textit{for the velocity field} at the high mode-index limit (e.g., the 1000th mode), which is characterized by the finest spatial scales. This leads to an energy overshoot in the $T_{100}$-generated PIV spectra in attached flames (Fig.~\ref{fig:pod_spectra}(c)) at the highest modes, and a correction for the $T_{10}$-generated model's undershoot in the detached flames at the highest modes (Fig.~\ref{fig:pod_spectra}(c)).

While there is room for improvement, the statistical trends of synthetic flames resemble those seen in the real data. POD spectra reveal that spatial statistical trends are largely insensitive to the generated slab size at larger scales, with most deviation between $T_{10}$ and $T_{100}$ models observed at the higher mode indices (smaller scales). Combined with the consistency of temporal trends observed in the $T_{100}$ model, this points to the promising potential for spatiotemporal ML-based generation of turbulent flames across fundamentally distinct regimes. Ultimately, the above analysis serves as a form of verification for the conditional, spatiotemporal generation of experimental-data based flames: the x-prediction, pixel-based transformer approach to diffusion modeling is a viable and powerful strategy to this end.

\subsection{Synthesizing Flame Transitions}
\label{subsec:transition_synthesis}

Section~\ref{subsec:val} demonstrated how the diffusion model can be used to generate statistically consistent spatiotemporal flame dynamics in both attached and detached flame configurations. To further demonstrate model capability, the objective of the present section is to leverage the \textit{same} model in an extrapolative context, namely, to use the conditional diffusion model to synthesize flame transitions unseen during training. Transitions in both directions (liftoff and reattachment) will be targeted. This is motivated by the broader objective of data augmentation in data-poor environments, particularly for generating experimental data corresponding to combustion extreme events, for which the aforementioned flame transitions are practical examples \cite{malik_pecs}.

\subsubsection{Denoising Model for Transitioning Flames}

Recall that in the inference stage, the generation of spatiotemporal data requires integrating the ODE in Eq.~\ref{eq:ode}, which is repeated below for convenience:
\begin{equation*}
    \frac{d \bm{z}_s (\tau)}{d \tau}  = \hat{\bm {v}}_s({\bm z}_s, \tau, y_s).
\end{equation*}
The generation process is tied to the model for the right-hand side (the modeled velocity), which comes from the x-prediction setup described in Sec.~\ref{sec:methods}. To generate attached flame dynamics, a fixed $y_s=0$ (the attached class label) is applied during integration; for lifted flame dynamics, a fixed $y_s=1$ (the detached class label) is applied.

It is emphasized that $\hat{\bm v}_s$ governs the dynamics in the $CTHW$-dimensional phase space along a pseudo-time axis, prescribing the evolution of a noised slab into the desired denoised slab (here, attached or lifted flame slabs), as illustrated in Fig.~\ref{fig:Denoise}. The transformer is therefore trained to push the pure noised slab onto the respective data manifolds for attached and lifted flame dynamics, and indeed accomplishes this task, as shown throughout Sec.~\ref{subsec:val}. As such, the task of transition dynamics generation can then be cast as formulating a new right-hand side (a new modeled velocity) for the above ODE, such that (a) invocation of the new modeled velocity results in a procedure that turns a pure noise time-slab into a transitioning flame, instead of attached or detached flames, (b) the way in which a transitioning flame is generated does not require a separate training stage, and instead requires only the model trained with knowledge of attached and detached flame dynamics in isolation, and (c) the design of the new transitioning modeled velocity allows for control of the direction of transition (liftoff versus reattachment). 

This is accomplished here through the following simplifying assumption: \textit{the modeled velocity for transition is cast as a weighted linear combination of the modeled velocities for attached and detached flames.} This translates to a representation of the transitioning flame as a superposition of the dynamics of attached and detached flames. Through this assumption, the modeled velocity for the denoising of a transitioning flame can be defined as
\begin{equation}
    \label{eq:transition_velocity}
    \hat{\bm v}_{s,\text{tra}}({\bm z}_s, \tau) = (1 - {\bm w})\odot \hat{\bm v}_{s,\text{att}}({\bm z}_s, \tau) + {\bm w} \odot \hat{\bm v}_{s, \text{det}}({\bm z}_s, \tau),
\end{equation}
where $\hat{\bm v}_{s,\text{tra}}({\bm z}_s, \tau) \in \mathbb{R}^{C \times T \times H \times W}$ is the modeled velocity for a transitioning flame slab, $\hat{\bm v}_{s, \text{att}}=\hat{\bm v}({\bm z}_s, \tau, 0) \in \mathbb{R}^{C \times T \times H \times W}$ is the modeled velocity for an attached flame slab, $\hat{\bm v}_{s, \text{att}}=\hat{\bm v}({\bm z}_s, \tau, 1) \in \mathbb{R}^{C \times T \times H \times W}$ is the same for detached, and ${\bm w} \in \mathbb{R}^T$ is a temporal weighting vector whose values lie within the range of $[0,1]$. The $\odot$ symbol in Eq.~\ref{eq:transition_velocity} denotes element-wise multiplication along the physical time axis (the $T$ frames). Therefore, specification of ${\bm w}$ provides control of the transition direction -- a ${\bm w}$ with values starting at 0 and ending at 1 encodes a flame liftoff, and vice-versa for reattachment. Crucially, alongside the direction, the design of ${\bm w}$ also provides a mechanism for controlling the \textit{rate} of the generated transition -- a sudden jump from $0$ to $1$ encodes a fast transition, while a smooth ramp-up from $0$ to $1$ encodes a slow transition. 

\begin{figure}
    \centering

    \includegraphics[width=0.4\textwidth]{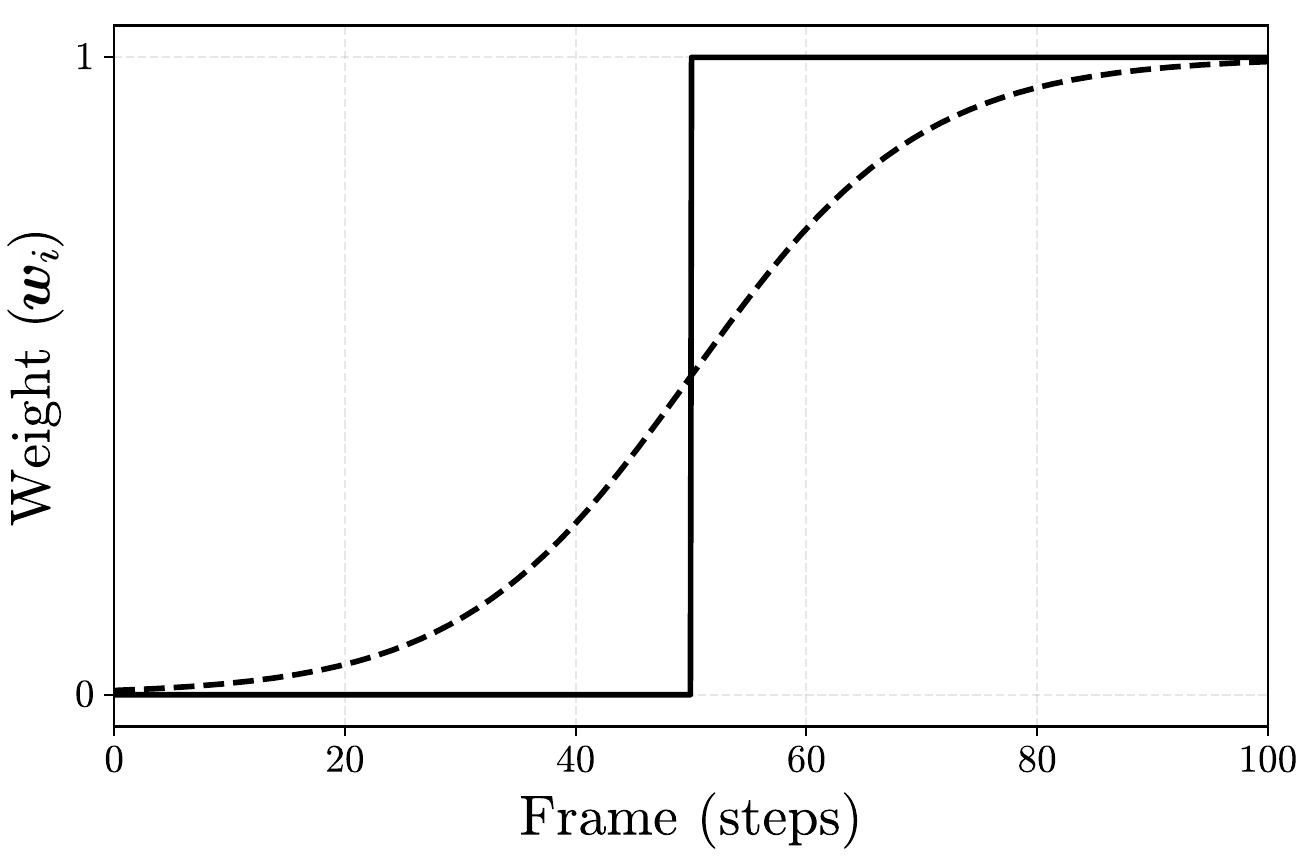}

    \caption{Temporal weight vectors ${\bm w}$ used in the modeled velocity (Eq.~\ref{eq:transition_velocity}) for synthesizing transitioning flames using the conditional diffusion model via Alg.~\ref{alg:transition}. Curves follow Eq.~\ref{eq:sigmoid}: solid is $\kappa=\epsilon$, dashed is $\kappa=10$.}
    \label{fig:sigmoid}
\end{figure}

For the liftoff transition (attached to detached), considering only the $T_{100}$ model (100 generated frames), the weight ${\bm w}\in \mathbb{R}^{100}$ is set here as the squeezed sigmoid centered at the 50th frame
\begin{equation}
    \label{eq:sigmoid}
    {\bm w}_i = \frac{1}{1 + \exp\left(-\frac{i-50}{\kappa}\right)}, \quad i=1,\ldots, 100,
\end{equation}
where $\kappa$ is the squeezing parameter that serves as a synthetic transition timescale (in frames). In the results below, synthesized transitions for (a) $\kappa=\epsilon$ (with $\epsilon$ denoting an arbitrarily small value much less than 1, but greater than 0, producing a step transition), and (b) $\kappa = 10$ (smooth transition) are shown for both liftoff and reattachment. These weight vectors are shown in Fig.~\ref{fig:sigmoid}. For generating reattachment transitions (detached to attached), the weight curves in Fig.~\ref{fig:sigmoid} are simply reflected along the x-axis.

The transition generation algorithm is provided in Alg.~\ref{alg:transition}, which is a direct extension of Alg.~\ref{alg:inference}. As with Alg.~\ref{alg:inference}, an Euler update is shown in Alg.~\ref{alg:transition} for simplicity -- for the results below, a Heun integrator is used. Before proceeding, it is noted that the transition-generation strategy is equivalent to a class-blending approach \cite{voleti2022mcvdmaskedconditionalvideo}, with an additional dependency along the physical time axis through the vector-based weighting function. Further, the transition generation approach used here is fundamentally different from (and extends) the cluster-based interpolation approach of Ref.~\cite{CARREON2023100238} used to synthesize instantaneous snapshots in transitional states, since the current approach is spatiotemporal dynamics-based, centers on a diffusion modeling framework, and provides control over the transition dynamics' directionality and timescale through specification of ${\bm w}$.

\begin{algorithm}[ht]
\caption{Conditional diffusion model inference for transition synthesis}
\label{alg:transition}
\begin{algorithmic}[1]
\Require Trained neural network $f$, attached/detached labels $y_s = 0, 1$, integration steps $N$, temporal weighting vector $\bm{w} \in \mathbb{R}^T$ (Eq.~\ref{eq:sigmoid})
\State Initialize noise slab: $\bm{z}_s(\tau_0) \sim \mathcal{N}(0, \mathbf{I})$
\State Set pseudo-time grid: $\{\tau_k\}_{k=0}^{N} \leftarrow \mathrm{linspace}(0, 1, N+1)$, \;\; $\Delta\tau \leftarrow 1/N$
\For{$k = 0, 1, \ldots, N-1$}
    \State Evaluate neural network for attached class: $\hat{\bm{x}}_{s,\text{att}} \leftarrow f(\bm{z}_s(\tau_k),\, \tau_k,\, 0;\theta_f)$
    \State Evaluate neural network for detached class: $\hat{\bm{x}}_{s,\text{det}} \leftarrow f(\bm{z}_s(\tau_k),\, \tau_k,\, 1;\theta_f)$
    \State Evaluate modeled velocities $\hat{\bm{v}}_{s,\text{att}}$, $\hat{\bm{v}}_{s,\text{det}} \leftarrow$ Eq.~\ref{eq:vpred}
    \State Evaluate transition velocity: $\hat{\bm{v}}_{s,\text{tra}} \leftarrow $ {Eq.~\ref{eq:transition_velocity}}
    \State Euler update: $\bm{z}_s(\tau_{k+1}) \leftarrow \bm{z}_s(\tau_k) + \Delta\tau\,\hat{\bm{v}}_{s,\text{tra}}$
\EndFor
\State \Return generated transitioning slab $\bm{x}_s \leftarrow \bm{z}_s(\tau_N)$
\end{algorithmic}
\end{algorithm}

\subsubsection{Evaluation of Generated Transitions}

Using the procedure described above, the following evaluates synthetically generated flame transitions in both directions (liftoff and reattachment), with the objective of understanding the degree to which characteristic instantaneous flow features during the transition process are captured in the synthetic data. To this end, real transition sequences are used as qualitative comparison baselines and are extracted from corresponding held-out transition sequences as described in Sec.~\ref{sec:data} (indicated by red regions in Fig.~\ref{fig:combustor_dataset}(c)). The held-out slabs containing the real transitions contain 300 frames, and generated transitions contain 100 frames, as they are constructed with the $T_{100}$ model. Although the generated slab size is different from the real counterparts, the objective here is to demonstrate the model's ability to synthesize transitions that capture consistently evolving flow features at specific snapshots during the transition process.

\begin{figure}[t]
    \centering
    \includegraphics[width=\textwidth]{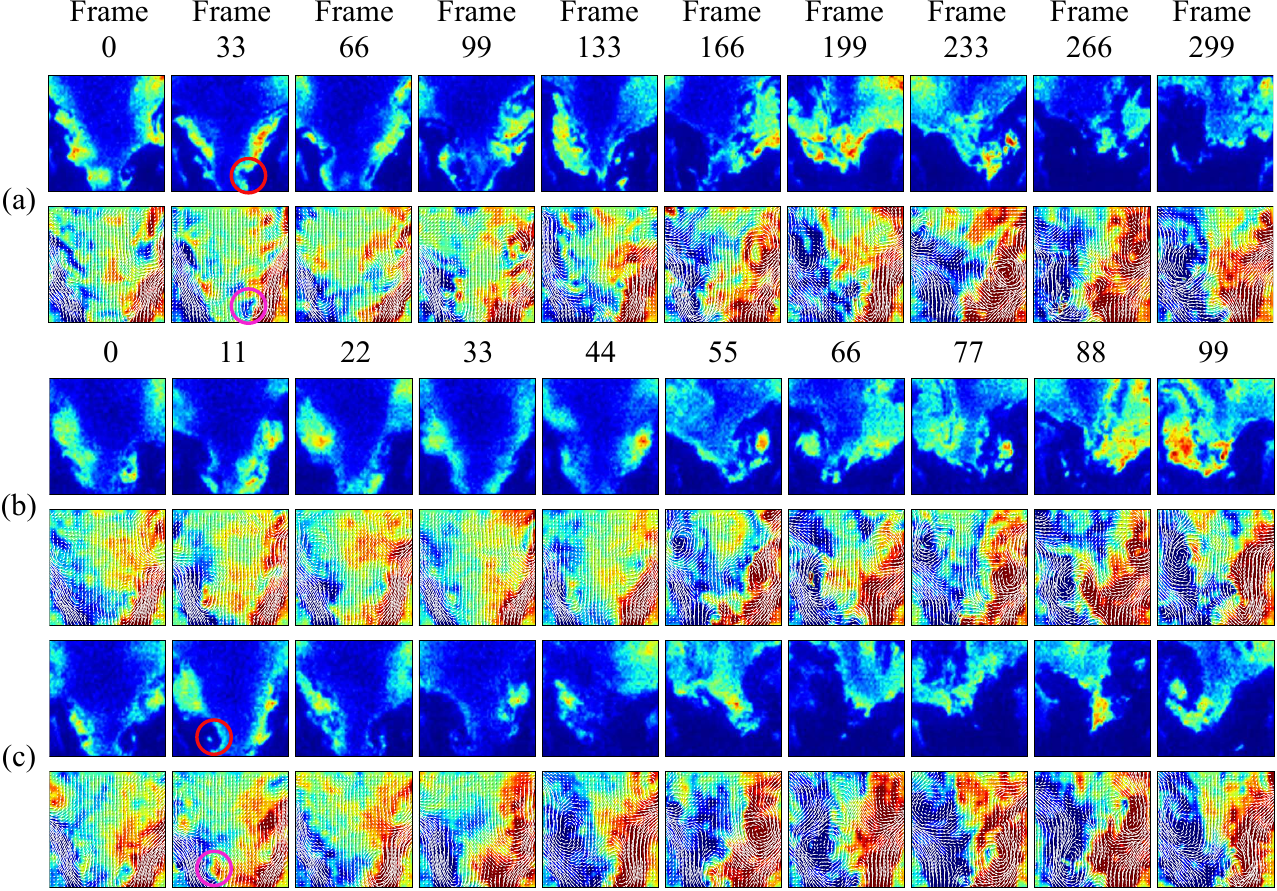}
    \caption{The flame liftoff process (attached to detached) sourced from (a) real, (b) synthetic ($\kappa = \epsilon$, step-based weight), and (c) synthetic ($\kappa=10$, smooth sigmoid weight) time slabs. Weight vectors used to generate (b) and (c) correspond to solid and dashed curves respectively in Fig.~\ref{fig:sigmoid}. Each sequence denotes the frame number in the respective slab, OH-PLIF field, and simultaneous PIV-Z field (with velocity vector overlays). Synthetic transitions are produced using the $T_{100}$ diffusion model (see Table~\ref{table:models}). Colorbars same as Fig.~\ref{fig:RVG-A}.}
    \label{fig:transition1}
\end{figure}

Ten evenly distributed snapshots for real and synthetic flame transitions for the liftoff process (attached-to-detached) are shown in Fig.~\ref{fig:transition1}, using the weight values shown in Fig.~\ref{fig:sigmoid} for synthetic transitions. Immediately evident is the fact that (1) the linear weighting procedure of Eq.~\ref{eq:transition_velocity}, combined with Alg.~\ref{alg:transition}, is capable of generating completely synthetic starting and ending flames in the desired attached and detached configurations \textit{within the same generated slab}, and (2) the transient progression from attached into the detached regime is more rapid in Fig.~\ref{fig:transition1}(b) (step weight) than in Fig.~\ref{fig:transition1}(c) (smooth weight), illustrating the fact that a synthetic transition rate can indeed be controlled through specification of the weighting function.

A key characteristic of flame liftoff in this configuration is the emergence of flame asymmetry, which signals the transition from the attached flame regime into the lifted regime through the development of the precessing vortex core \cite{AN2019267,shivam_crom_ctm}. This asymmetry is observed in the real liftoff sequence in the OH-PLIF field at (Fig.~\ref{fig:transition1}(a), Frame 133), which results in a progression to the formation of a hook-like flame structure in detached flames (Frames 199, 233), leading to complete liftoff. A qualitatively similar emergence of flame asymmetry is observed in the smooth synthetic transition in Fig.~\ref{fig:transition1}(c) in the 20-frame span roughly between frames 44-66, with frames 55 and 66 in particular mirroring the structure of OH-PLIF asymmetry emergence in the real flame. Interestingly, and consistent with the specification of the step-based weighting function, these structures are not observed in the plotted frames for Fig.~\ref{fig:transition1}(b), which instead depicts a direct jump to a detached frame within the 10-frame sequence from Frame 44 to 55. Inspection of corresponding velocity fields shows that the synthetic velocity data mirrors the trend in the real velocity data: as the flame lifts off, alternating recirculation zones across the centerline emerge, and the anti-symmetric PIV-Z velocity streaks spanning the full vertical length of the domains become dominant. 

The presence of local extinction zones in the attached regime (indicated by the red circles in Fig.~\ref{fig:transition1}), coincident with smaller localized recirculation zones (purple circles in PIV plots), is a key feature of the liftoff process \cite{AN2016228,AN2019267}. This feature is present in the real transition before more extreme flame asymmetry arises -- it is also observed in the synthetically-generated transition, but for the smooth-weight case, and \textit{not} the step-weight case. This finding points to a potential avenue for the application of the transition synthesis strategy as a means to isolate transition precursor features: the observation of this finer-scale extinction feature in Fig.~\ref{fig:transition1}(c), and not Fig.~\ref{fig:transition1}(b), can be attributed to design of its corresponding weight function (Fig.~\ref{fig:sigmoid}), which enforces predominantly -- \textit{but not completely} -- attached dynamics at that point in the frame sequence (specifically, Frame 22).

\begin{figure}[t]
    \centering
    \includegraphics[width=\textwidth]{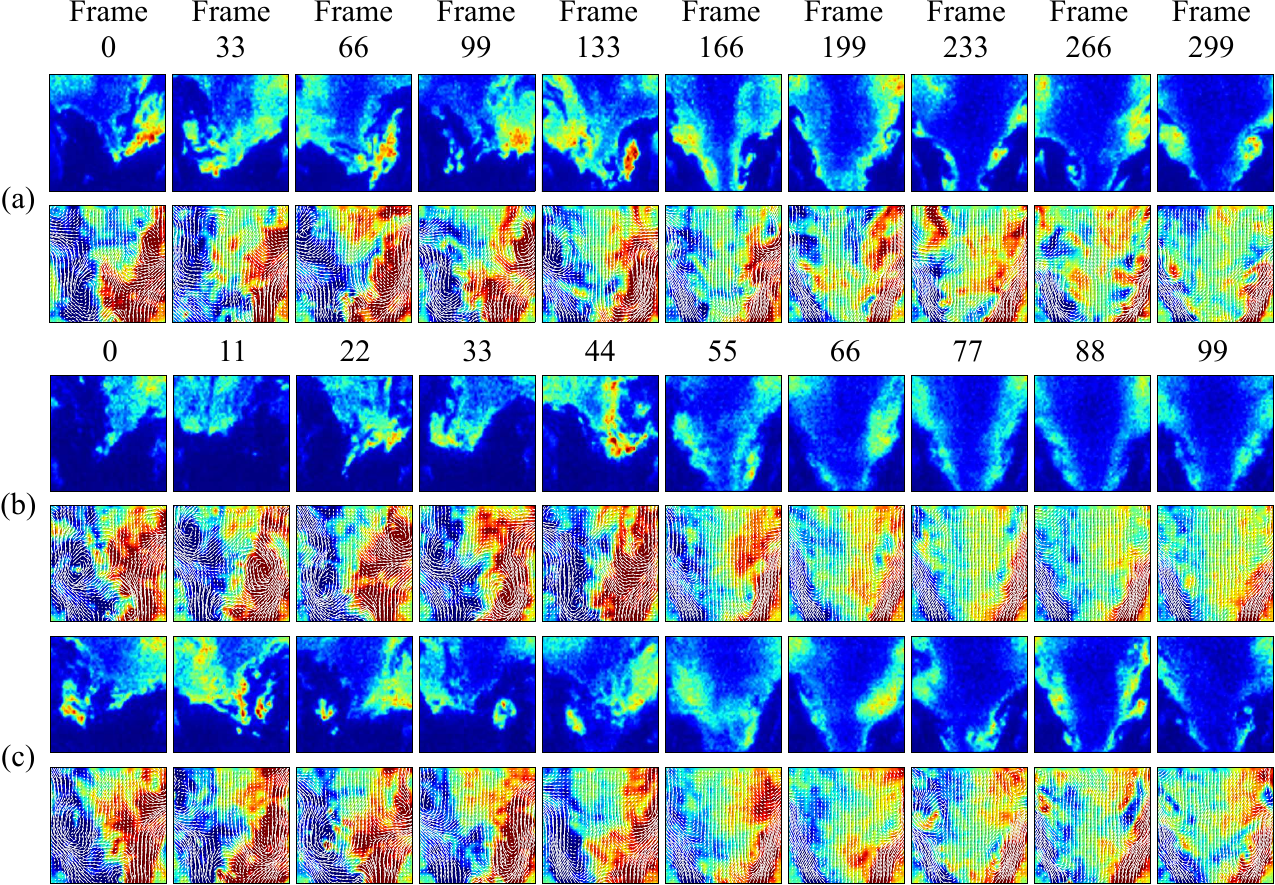}
    \caption{Same as Fig.~\ref{fig:transition1}, but for the reattachment transition (detached-to-attached).}
    \label{fig:transition2}
\end{figure}

Visualizations for the synthesized reattachment transitions are shown in Fig.~\ref{fig:transition2}. While the precursor mechanisms for the reattachment process are less clear \cite{shivam_crom_ctm}, trends analogous to the liftoff case are observed when comparing real and synthesized transitioning flames. Specifically, the overall progression of the lifted flame structure into an asymmetric attached flame, and then into a symmetric attached flame, is directly apparent through OH-PLIF visualizations, and the rate of this reversion is again shown to be controllable through the modification of the weight function. Comparisons of velocity frames also show alignment with features observed in the real reattachment process -- namely, the disappearance of the zig-zag distribution of recirculation zones as the flame re-attaches, and the reduction in the size of out-of-plane velocity streaks that correlate with an absence of the PVC structure.

\begin{figure}
    \centering

    \includegraphics[width=\textwidth]{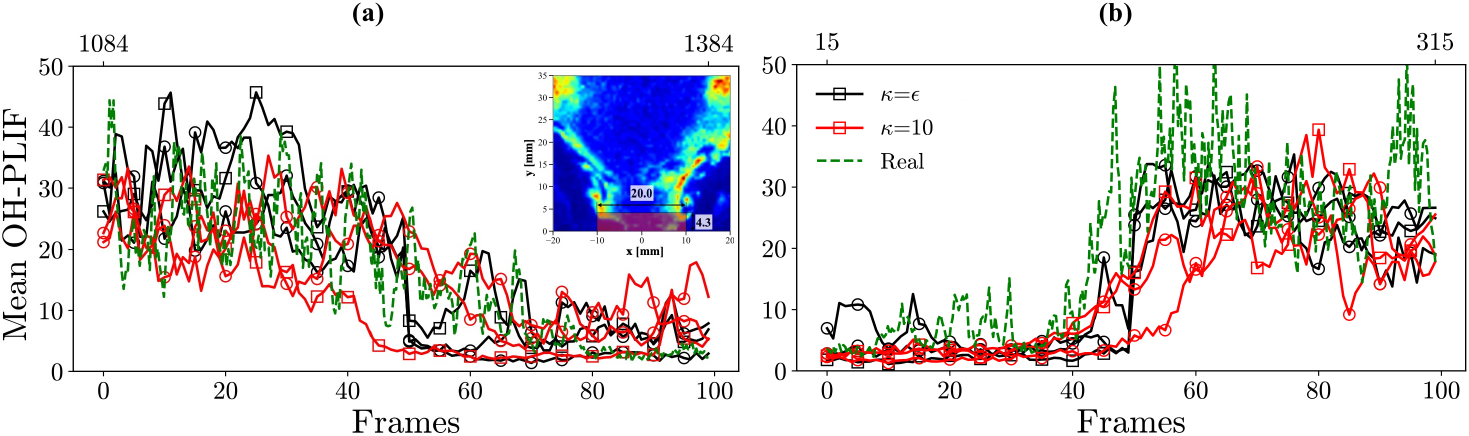}

    \caption{Temporal evolution of mean OH-PLIF concentration for (a) the liftoff process (attached to detached) and (b) the reattachment process (detached to attached), with mean OH values extracted from a box (20 mm $\times$ 4.3 mm) near the burner exit, as indicated by the inset. In each plot, black and red curves indicate $\kappa=\epsilon$ (step) and $\kappa=10$ (smooth) weighting for transition generation, respectively. Square markers indicate transition trajectories shown in Figs.~\ref{fig:transition1} and~\ref{fig:transition2}; circle markers indicate other sampled trajectories. Dashed green lines represent real transition sequences, corresponding to those shown in the same figures. The primary $x$-axis (lower labels) is used for generated transitions produced by the $T_{100}$ model (spanning 100 frames); the secondary $x$-axis (upper labels) is used for the real transition sequence (spanning 300 frames).}     
    \label{fig:transition_plif}
\end{figure}

Evolution of mean OH values collected from a rectangular sub-domain near the burner exit is provided in Fig.~\ref{fig:transition_plif}, which provides a more quantitative depiction of the transition process. Specifically, mean OH evolution from three synthetic transitions is shown for both liftoff (Fig.~\ref{fig:transition_plif}(a)) and reattachment (Fig.~\ref{fig:transition_plif}(b)), with values obtained from a real transition overlaid on each. Inspection of the mean OH curves shows how the synthetic transitions are largely a reflection of the weighting functions used to generate them: the $\kappa=\epsilon$ (step) weight results in a more pronounced, stark change in mean OH at the 50-frame mark, whereas the $\kappa=10$ weight shows a smoother change, highlighting the control offered by the weighting function in the transition generation model. Critically, even for a fixed weighting function, Fig.~\ref{fig:transition_plif} shows that the model produces \textit{diverse} transition samples -- the model is able to synthesize completely new transitions that are spatiotemporally coherent, while retaining relatively strong control over the generated transition timescale through the weight function. 

Comparison with real transition sequences, denoted by the green curves in Fig.~\ref{fig:transition_plif}, shows that the synthesized trajectory is able to capture the overall evolution trend of transition trajectories in both directions. However, some difficulty is faced in post-transition regimes: in Fig.~\ref{fig:transition_plif}(a), some realizations are not as cleanly lifted off (indicated by the high-variation mean OH signal at the end of the synthetic liftoffs), and in Fig.~\ref{fig:transition_plif}(b), the large range variation in mean OH signal post-transition is not captured in the synthetic flames.

Despite these deviations, the observed similarity between the real and synthetic transitioning flame features points to a promising ability of the generative model to synthesize flame transitions, a highly challenging extrapolative task for the diffusion model. It is emphasized that, while the model for the denoising transition velocity in Eq.~\ref{eq:transition_velocity} is a \textit{linear} combination of attached/detached denoising velocities, the resulting synthetic transition dynamics are strongly \textit{nonlinear}, as evidenced by Fig.~\ref{fig:transition_plif}. These findings ultimately warrant more detailed investigation into the denoising velocity model for transition in future work, for both the present application and others in which flow transition phenomena are of high engineering relevance but remain data-sparse.

\section{Conclusion}
\label{sec:conclusion}

In this work, a conditional diffusion model was developed to generate synthetic, simultaneous experimental laser diagnostic data (specifically, OH-PLIF and multi-component PIV measurements) for turbulent flame dynamics in a swirl-stabilized combustor configuration \cite{AN2016228,AN2019267}, building on previous work \cite{CARREON2023100238}. This was accomplished using a flow matching framework following the ``x-prediction'' variant, wherein the predicted denoising velocity is recovered using a linear noise schedule assumption combined with a neural network-based model for the denoising velocity \cite{jit}. For the neural network, a spatiotemporal transformer was used, designed to operate directly on the pixels composing space-time slabs of evolving flames, eliminating the need to predefine a latent space dimensionality for the denoising process. With this approach, in the inference stage, trained diffusion models were made capable of generating entire flame trajectories at once, in both attached and detached flame regimes, using a class-conditioning strategy \cite{dit}. This resulted in a generative model capable of synthesizing new, spatiotemporally coherent experimental data-based trajectories for complex turbulent flames, with each frame in the generated trajectories comprised of simultaneous OH-PLIF and PIV fields (four total measurements per pixel).

Demonstrations were carried out using two models: (1) the $T_{10}$ model, trained to generate spatiotemporal slabs with a 10-frame time span, and (2) the $T_{100}$ model, trained to generate slabs with a 100-frame time span, allowing for assessment of the effect of increased dynamical range on generation quality. Analysis of flame dynamics in physical space, coupled with detailed assessment of statistical consistency via proper orthogonal decomposition, revealed that synthetic flames produced by the generative models retain all major physical features in both attached and lifted regimes. This confirmed that the approach successfully allows for conditional generation of simultaneous OH-PLIF/PIV data in the markedly different regimes of attached and detached flames. It was found that, under the transformer architecture used here, while both $T_{10}$ and $T_{100}$ models behaved similarly overall in terms of spatial statistical consistency, the $T_{100}$ model generated slight deviations in predicted energy content at smallest spatial scales relative to the $T_{10}$ model, at the tradeoff of the $T_{100}$ model providing the necessary temporal range to capture the characteristic dynamics of the detached flame regime (e.g., the precessing vortex core).

Additionally, motivated by the broader objective of data augmentation in extreme event-prone combustion environments \cite{malik_pecs}, an extrapolation task consisting of \textit{transition synthesis} was carried out. Here, the generative model was tasked with synthesizing new flame transitions corresponding to liftoff (evolution of the flame from an attached to detached state) and reattachment (detached to attached), both of which are highly transient processes unobserved during training. Transition generation was achieved by constructing a model for the transition denoising velocity as a linear combination of attached and detached denoising velocities, using a time-evolving weight function. This formulation resulted in synthetic transitions exhibiting qualitatively similar flow features to real held-out transitions, resulting in a generalizable methodology that (a) allows for the control of generated transition directions and timescales, and (b) retains sample-to-sample variability in generated transitions in the process. These qualities lead to a promising pathway for the utilization of generative models as a new means for data exploration in data-sparse environments, complementing both experiments and computational fluid dynamics-based approaches.

Key challenges in the generation process remain, and there are many avenues for future work. First, while the pixel-based transformer approach used here eliminates the need to prescribe a latent dimensionality during generation, it relies on a scaled dot-product attention operation that is quadratic in complexity with respect to the number of patches per spatiotemporal slab, leading to relatively large training and inference costs for large slabs. Alternative attention model forms \cite{swin_v2}, distillation strategies \cite{diffusion_distillation}, and distributed attention methods \cite{shyam_distributed,aeris} are promising directions to alleviate this cost. Further, from the optimization standpoint, incorporation of spectral losses in the flow matching formulation may help improve statistical consistency to an even greater degree at training time \cite{dj_spectral}, with diffusion posterior sampling \cite{xu_dps_neurips} and more robust integration methods \cite{dpm_solver} capable of providing improvements at inference time. Exploration of these and other directions is left for future work.

\section{Acknowledgements}
The authors thank Prof. Adam Steinberg (Georgia Institute of Technology) for sharing the experimental data used in this work. Discussions with Shyam Sankaran on the topic of diffusion modeling are gratefully acknowledged. This research used resources of the Argonne Leadership Computing Facility (ALCF) at Argonne National Laboratory, which is a U.S. Department of Energy Office of Science User Facility operated under Contract No. DEAC02-06CH11357.

\bibliographystyle{elsarticle-num}
\biboptions{sort&compress}
\bibliography{references}

\end{document}